\documentclass{article}

\usepackage{arxiv}
\usepackage{multirow}
\usepackage[utf8]{inputenc} 
\usepackage[T1]{fontenc}    
\usepackage{hyperref}       
\usepackage{url}            
\usepackage{booktabs}       
\usepackage{nicefrac}       
\usepackage{microtype}      
\usepackage{lipsum}
\usepackage{amsmath,amsthm,amssymb}
\usepackage{amsfonts}
\usepackage{mathdots}
\usepackage{bm}
\usepackage{yhmath}
\usepackage{mathrsfs}
 \usepackage{resizegather}
\usepackage{enumitem}
\usepackage{caption}
\usepackage{subcaption}
 \usepackage{float}
\usepackage[dvips]{graphicx}
\usepackage{xcolor}
\usepackage{mathtools}
\newcommand{\abs}[1]{\left\lvert#1\right\rvert}

\title{Product Partition Dynamic Generalized Linear Models}
\newcommand\numberthis{\addtocounter{equation}{1}\tag{\theequation}}

\author{
Victor S.~Comitti\\
  Instituto Federal Sudeste de Minas Gerais\\
  Campus Bom Sucesso\\
   \texttt{victor.comitti@ifsudestemg.edu.br} \\
    \And
Fábio N.~Demarqui \\
 Departamento de Estatística\\
  Universidade Federal de Minas Gerais\\
  \texttt{fndemarqui@est.ufmg.br} \\
   \And
 Thiago R.~dos Santos \\
 Departamento de Estatística\\
  Universidade Federal de Minas Gerais\\
  \texttt{thiagords@est.ufmg.br } \\
  \And
  Jéssica da Assunção ~Almeida \\
 Departamento de Estatística\\
  Universidade Federal de Minas Gerais\\
  \texttt{ jessica.assuncao92@hotmail.com} \\
}

\begin{document}
\maketitle

\begin{abstract}
Detection and modeling of change-points in time-series can be considerably challenging. In this paper we approach this problem by incorporating the class of Dynamic Generalized Linear Models (DGLM) into the well know class of Product Partition Models (PPM). This new methodology, that we call DGLM-PPM, extends the PPM to distributions within the Exponential Family while also retaining the flexibility  of the DGLM class. It also provides a framework for Bayesian multiple change-point detection in dynamic regression models. Inference on the DGLM-PPM follow the steps of evolution and updating of the DGLM class. A Gibbs Sampler scheme with an Adaptive Rejection Metropolis Sampling (ARMS) step appended is used to compute posterior estimates of the relevant quantities.  A simulation study shows that the proposed model provides reasonable estimates of the dynamic parameters and also assigns high change-point probabilities to the  breaks introduced in the artificial data generated for this work. We also present a real life data example that highlights the superiority of the DGLM-PPM over the conventional DGLM in both in-sample and out-of-sample goodness of fit measures.
\end{abstract}


\section{Introduction}

Change-points, or structural breaks, occur quite often when one is trying to model a time series. This type of phenomenon can be seen as a result of some external intervention that changes some driving parameters of the process under analysis -- usually the mean or the variance (or even both, in some cases). Change-points can appear one or multiple times in a time series,  failing to model them correctly can lead to large forecasting errors. For this reason, intervention analysis plays a crucial role in many fields such as economics, finance, engineering, climatology, hydrology, among others.


The literature on change-points is vast and covers both Bayesian and classical approaches. Under the classical framework, most of the single change-point detection models are variations of the popular Cumulative Sum (CUSUM) process. These models are usually formulated in terms of hypothesis tests, with the null hypothesis being the stability of the parameters. An extensive overview of these methods can be found in Csorgo \& Horváth (1997) ~\cite{csorgo97}, Perron (2006) ~\cite{perron2006} and Aue (2013) ~\cite{aue2013}. Tests that allow for the detection of multiple unknown change-points are, in most cases, least-squares type. Important works on this setting include Bai \& Perron (1998) ~\cite{bai98}, Bai (1999) ~\cite{bai99}, Qu \& Perron (2007)~\cite{qu2007}, Kurozumi \& Tuvaandorj (2011) ~\cite{kurozumi2011} and Preuss, Puchstein and Dette (2015) ~\cite{preuss2015}.


In this paper, we are concerned with identifying and modeling multiple change-points in a Bayesian framework. Bayesian modeling of structural breaks in time series generally considers that the underlying process is governed by a latent discrete state vector in a state space model, usually taken as piecewise constant, following a Markovian evolution. Under this perspective,  temporal heterogeneities can be interpreted as an abrupt change in the state variable driving the process. Point and interval estimates for the state parameters can be obtained using conventional Bayesian computational tools such as Monte Carlo Markov Chain (MCMC) methods. Important papers following this type of approach include Chib (1998) ~\cite{chib98}, Lai (2005) ~\cite{lai2005}, Lai \& Xing (2011) ~\cite{lai2011} and Martínez \& Mena (2014) ~\cite{martinez2014}. More recently, methods based on particle filters are also receiving attention. For works using this setting, we refer the reader to Caron, Doucet  \& Gottardo (2012) ~\cite{caron2012} and da Silva and da Silva (2017) ~\cite{dasilva2017}.


Another Bayesian approach to the change-point problem that has drawn considerable attention over the last decades is the class of Product Partition Models (PPM) proposed by Hartigan (1990) ~\cite {hartigan90} and posteriorly extended by Barry \& Hartigan (1992, 1993). Models within this class induce a block structure that subdivides a given data set into contiguous blocks (or components) of similar observations forming a partition that we will denote here by $\rho$. It is assumed that these observations  are conditionally independent given a vector of parameters $\bm{\theta}_{\rho}^{(j)}$, where $j \in \{1, 2, 3, \cdots\ b\}$ indexes every block in the partition and $b$ denotes the cardinality of $\rho$. For every two components $i$ and $j$ from $\rho$ such that $i\neq j$, it holds that $\bm{\theta}_{\rho}^{(i)}\neq \bm{\theta}_{\rho}^{(j)}$ -- that is, the observations in different blocks are subject to different underlying processes. Posterior estimates for the number of change-points (or blocks) can be obtained exactly or approximately using Markov sampling methods. Many works related to the Barry \& Hartigan proposal can be found in the literature. Loschi \& Cruz (2002) ~\cite{loschi2002}  study the influence that different  prior specifications for the degree of similarity between observation in the same block have on the PPM product estimates; Loschi \& Cruz (2005) ~\cite{loschi2005} and Fearnhead (2006) ~\cite{fearnhead2006}  provide a method for sampling direct from the posterior distribution of the number of change-points; Loschi, Pontel \& Cruz (2010) ~\cite{loschi2010} extend the PPM to detect multiple change-points in regression problems. For a comprehensive overview of the PPM literature we suggest Quintana, Loschi and Page (2018) ~\cite{quintana2018}.


In this work, we propose a new structure  that incorporates the Dynamic Generalized Linear Models (DGLM) class introduced by West, Harrison \& Migon (1985) ~\cite{WHM85} into the PPM. This new formulation, which from here on, we shall call DGLM-PPM, inherits the flexibility of the DGLM class, allowing for PPM regression models with dynamic structure and observations that belong to any distribution in the Exponential Family (EF). Our proposal also permits retrospective analysis, as in most change-point models, and online inference, which is a significant advantage concerning the existing literature on Bayesian methods for intervention analysis. Inferences on the DGLM-PPM class follow the sequential nature of the Bayesian approach with evolution and updating steps. We also make use of Bayesian conjugacy, thus providing an efficient path for obtaining a closed form expression for the predictive distribution (or data factor). Samples of the partitions are obtained with the help of the Gibbs Sampler scheme proposed by Barry \& Hartigan (1993) \cite{barry93}, and the optimal discount factor associated with the block structure of the model is estimated by appending an  Adaptative Rejection Metropolis Sampling (ARMS) step within the Gibbs algorithm. We also conduct a simulation study of the DGLM-PPM Poisson and present a real-life application with Poisson responses showing that our method outperforms the conventional DGLM.


This article is organized as follows: In Section 2, we present the PPM as described by Barry \& Hartigan (1992) ~\cite{barry92}. This formalism is extended in Section 3, where we introduce the DGLM-PPM class along with the Gibbs sampling scheme used in this work to obtain inference about the partition. In Section 4 we derive the DGLM-PPM Poisson as an example. The simulation experiment is presented in Section 5.  In Section 6  we apply the conventional DGLM and the DGLM-PPM to the well-known coal mining disaster time series and compare their performances. Finally, Section 7 is reserved for the final remarks.


\section{Product Partition Models}
\label{sec:PPM}

Barry \& Hartigan (1992) ~\cite{barry92} define a partition as a group of contiguous blocks.  It is assumed that, for every block in the partition, the observations obey a different probability model. More formally, the authors define a block as follows: let
$\mathbf{Y}=\{Y_{1},\cdots, Y_{n}\}$ be a sequence of consecutive observations, $I=\{0, 1,2,\cdots, n\}$ a set of indexes, $\rho=\{i_{0}, i_{1}, \cdots, i_{b}\}$ a random partition of the set $I$ such that $0=i_{0}<i_{1}<\cdots<i_{b}=n$, and $B$ a random variable denoting the number of blocks in a given partition. In the case where $B=b$, the partition can be written as

\[
[Y_{1},\cdots, Y_{i_{1}}]\text{,}[Y_{i_{1}+1},\cdots, Y_{i_{2}}]\text{,}\cdots \text{,}[Y_{i_{b-1}+1},\cdots, Y_{i_{b}}].
\]

In the expression above each block is denoted by $\mathbf{Y}_{[i_{j-1}i_{j}]}=[Y_{i_{j-1}+1}\cdots,Y_{i_{j}}]^{\intercal}$ with  $j=1,2,\cdots, b$. The $j$-th block can be identified as the set of observations given by $i+1, \cdots, j$, where $i,j \in \rho$ and $i<j$. Barry \& Hartigan (1992) define a cohesion function, $c_{\rho}^{(j)}$, that measures how likely the observations are to cocluster in the $j$-th component of the partition. One can also think of  cohesion as transition probabilities of the Markov Chain (MC) defined by the endpoints of each block ($i_{0}, i_{1}, \cdots,i_{b}$ ).



According to Loschi \& Cruz (2002) \cite{loschi2002} the random set $(Y_{1},\cdots,Y_{n};\rho)$ follows a PPM if the two conditions below are verified:

\begin{enumerate}

\item The prior distribution that describes the probability that a partition $\rho$ have endpoints $\{i_{0}, i_{1}, \cdots, i_{b}\}$ has a product form given by:
\begin{equation}
P(\rho=\{i_{0}, i_{1}, \cdots, i_{b}\})=\frac{1}{K}\prod_{j=1}^{b}c_{\rho}^{(j)},
\label{prho}
\end{equation}
where $K=\sum_{\mathcal{C}} \prod_{j=1}^{b}c_{\rho}^{(j)}$ is a normalizing factor with $\mathcal{C}$ representing  all possible partitions of the set $I$ into $b$ contiguous blocks with  endpoints satisfying $0=i_{0}<i_{1}<\cdots<i_{b}=n$, $\forall~b \in I$.

\item Conditionally on $\rho=\{i_{0}, i_{1}, \cdots, i_{b}\}$ the observations $Y_{1},\cdots,Y_{n}$ have  the following  joint distribution:

\begin{equation}
p(Y_{1},\cdots,Y_{n} \mid \rho=\{i_{0}, i_{1}, \cdots, i_{b}\})=\prod_{j=1}^{b}p_{j}(\bm{Y}_{\rho}^{(j)}),
\label{predgeral}
\end{equation}

where $p_{j}(\bm{Y}_{\rho}^{(j)})$ is the density of the random vector $\bm{Y}_{\rho}^{(j)}$.
\end{enumerate}

Under those two assumptions, according to  Barry \& Hartigan (1992), the posterior distribution of the partition $\rho$ will follow the same product form of the Equation (\ref{prho})  with the  posterior cohesion for the $j$-th block defined as

\begin{equation}
c_{\rho}^{*(j)}=c_{\rho}^{(j)}p_{j}(\bm{Y}_{\rho}^{(j)}).
\end{equation}

The whole construction presented  so far does not assume any parametric form for the PPM. In the parametric approach one considers that each observation $Y_{k}$, $k \in \{1,2,\cdots, n\}$ can be described by a marginal density conditioned on an unknown parameter $\theta_{k}$ that we will denote by $p(Y_{k}\mid \theta_{k})$. Given $\theta_{1},\cdots,\theta_{n}$, $Y_{1},\cdots,Y_{n}$ are assumed to be conditionally independent with joint density $\displaystyle \prod_{j=1}^{b}p(\bm{Y}_{\rho}^{(j)}\mid \theta_{\rho}^{(j)})$. It is also assumed that, within each block $j$, the corresponding observations are identically distributed. Particularly, given the partition $\rho=\{i_{0}, i_{1}, \cdots, i_{b}\}$, $b\in I$ and a block $[i_{r-1}i_{r}]$ we have $\theta_{k}=\theta_{[i_{r-1}i_{r}]}$, $\forall$ $k$ such that $i_{r-1}+1<k<i_{r}$ and $r\in I$; that is, within, say, the $j$-th block, it must hold that: $\theta_{\rho}^{(j)} = \theta_{i_{j}} = \theta_{i_{j}+1} = \cdots = \theta_{i_{j}+ n_{j}}$, where $n_{j}$ denotes the number of observations in block j. Also, for each block $j$, we assign a correspondent block prior distribution $p(\theta_{\rho}^{(j)})$. Thus, the block predictive function (or data factor) $p(\bm{Y}_{\rho}^{(j)})$ can be calculated from
\begin{equation}
p(\bm{Y}_{\rho}^{(j)})=\int_{\Theta_{\rho}^{(j)}}p(\bm{Y}_{\rho}^{(j)}\mid \theta_{\rho}^{(j)})p_{j}(\theta_{\rho}^{(j)})d\theta_{\rho}^{(j)},
\label{blockppred}
\end{equation}
and the block posterior density is given by,
\begin{equation*}
    p(\theta_{\rho}^{(j)}\mid \bm{Y}_{\rho}^{(j)})=\frac{p(\theta_{\rho}^{(j)})\prod_{i_{j-1}+1}^{i_{j}}p(Y_{k}\mid \theta_{\rho}^{(j)})}{p(\bm{Y}_{\rho}^{(j)})}, \end{equation*}
 for $j=1, \cdots, b$. Equation (\ref{blockppred}), in theory, can be solved by means of numerical integration techniques. In practice, however, this procedure can be very difficult or even computationally impossible depending on the dimensionality of the problem. This is why, for the parametric PPM, it is very important to explore Bayesian conjugacy whenever possible. Our proposal provides a straightforward path to achieve that and obtain closed form solutions for the block predictive distribution and for the moments of the block posterior densities.

 The posterior distribution of the partition $\rho=\{i_{0}, i_{1}, \cdots, i_{b}\}$ has a product form given by,
 \begin{equation}
P(\rho=\{i_{0}, i_{1}, \cdots, i_{b} \} \mid \mathbf{Y})=\frac{\prod_{j=1}^{b}c^{*(j)}_{\rho}}{\sum_{\mathcal{C}}\prod_{j=1}^{b}c^{*(j)}_{\rho}},
\label{postrho}
\end{equation}
where $c^{*(j)}_{\rho}=c_{\rho}^{(j)}p(\bm{Y}_{\rho}^{j})$ represents the posterior cohesion associated with the block $[i_{j-1}i_{j}]$.

Finally, the posterior distribution of $\theta_{k}$  can be written as
\begin{equation}
p(\theta_{k} \mid Y_{1},\cdots,Y_{n})=\sum_{i_{j-1}<k\leq i_{j}}r_{\rho}^{* (j)}p_{j}(\theta_{k} \mid \bm{Y}_{\rho}^{(j)})\text{,}
\end{equation}
where $r^{*(j)}$ denotes the posterior relevance defined as follows
\begin{equation*}
  r^{*(j)}_{\rho}=r^{*(j)}_{\rho}=\frac{\lambda_{i_{0}i_{j-1}}c_{\rho}^{*(j)}\lambda_{i_{j}i_{b}}}{\lambda_{i_{0}i_{b}}},
\label{relevance}
\end{equation*}
 with $\lambda_{i_{j-1}i_{j}}=\sum \prod_{k=1}^{b}c_{\rho}^{*(j)}$, the summation being over all sets $i=i_{0}<i_{1}<\cdots<i_{b-1}<i_{b}=j$. This quantity represents the probability that the $j$-th block belongs to the partition $\rho$ given the data $\bm{Y}$. The posterior expected value of $\theta_{k}$, or product estimate,  can be computed from

\begin{equation}
\mathbb{E}(\theta_{k}\mid Y_{1},\cdots,Y_{n})=\sum_{i_{j-1}<k\leq i_{j}}r^{*(j)}_{\rho}E(\theta_{k} \mid \bm{Y}_{\rho}^{*(j)}),
\label{esperancapost}
\end{equation}

 Thus, the PPM provides a framework in which inferences on clustered parameters can be obtained for each block $j$ using standard Bayesian tools and the observations within the blocks. The posterior distributions for every $\theta_{k}$, $k=1,\cdots,n$ are computed, according to Equation (\ref{esperancapost}), as weighted averages over the densities associated with the blocks containing $\theta_{k}$, where the weights are taken as the posterior relevances $r^{*(j)}_{\rho}$ defined in Equation (\ref{relevance}).

An explicit calculation of the posterior relevances in Equation (\ref{esperancapost}) is possible as long as the exact posterior cohesions are available. However, since the possible number of partitions grows exponentially as the number of observations increase, it can be too expensive from a computational point of view,  specially for large samples. In Section 3 we show that the Gibbs Sampling approach introduced by Loschi \& Cruz (2002) can be used to overcome this problem and find posterior estimates of $B$ and $\rho$ at a reasonable computational cost. We also propose an adaptation of this method that is more appropriate for computing the parameters of the DGLM-PPM.

\section{DGLM-PPM}

An important aspect concerning the PPM class is that the parameters $(\theta_{\rho}^{(1)} \cdots \theta_{\rho}^{(b)})$, with $b\in I$, are allowed to be time-varying as long as: i) given $\theta_{1},\cdots,\theta_{n}$, the observations $(Y_{1},\cdots,Y_{n})$ are conditionally independent; ii) The joint distribution of the observations and partitions have a product form. Thus, assuming that those two conditions are met, a PPM can be used to detect change-points in a time series. In this section, we introduce  the DGLM-PPM -- a new class of dynamic models that incorporates the DGLM of West, Harrison \& Migon (1985) ~\cite{WHM85} into the PPM class. Under this new framework,  the two conditions that define a PPM are preserved and the parameters $\theta_{\rho}^{(1)},\cdots, \theta_{\rho}^{(j)}, \cdots, \theta_{\rho}^{b}$ associated to each block $j$ in the partition $\rho$ are, now, allowed to be dependent. This new formulation is very rich since it retains the flexibility of the DGLM while also permitting the detection of multiple change-points using the block structure of the PPM.

Another approach to the change-point problem that also considers across cluster correlation can be found in  Ferreira, Loschi \& Costa (2014). In this work the authors assume that the observations follow a Normal distribution with cluster mean, $\mu_{\rho}^{(j)}$, and variance $\sigma_{\mu}^{2}$. A Markovian dependence in the block structure  is introduced by the prior  $(\mu_\rho^{(j)} \mid \mu_{\rho}^{(j-1)}, \sigma^{2}) \sim N(\mu_{\rho}^{(j-1)}, \sigma^{2}_{\mu})$ and, as a consequence, the inference of the model requires the solution of a high dimensional multiple integral involving the mean, $\mu_{\rho}^{(j)}$, of each cluster observed in the partition prior to observation $i_{j}+1$.  In this Section we show that the DGLM-PPM provides a natural path for avoiding high dimensional integral just by using properties of the EF and the inference structure of the DGLM class. Also, our model is not restricted to normality or Gaussian assumptions, making it much more general than other works in the literature.


\subsection{Basic Structure of the model}

Consider the quantities $I$, $\rho$ and $B$ defined in the previous section and let $\bm{Y}=(y_{1},\cdots,y_{n})^{\intercal}$ be a time series with observations derived from a distribution belonging to the uniparametric EF. If $B=b$, the partition is composed of $b$ blocks, each denoted by: $Y_{\rho}^{(j)}=(y_{i_{j-1}+1}\cdots,y_{i_{j}})^{\intercal}\equiv (y_{j1},\cdots, y_{j_{nj}})^{\intercal}$, for $j=1,2,\cdots,b$, where $y_{jk}$ represents the $k$-th element and $n_{j}$ is the number of observation in the j-th block. The DGLM-PPM is defined by the set of equations described below.

\begin{itemize}
    \item Observation equation:
    \begin{equation}
p(y_{jk}\mid \eta_{jk}, \tau_{t}, \rho)=\exp[\tau_{jk}(y_{jk}\eta_{jk}-a(\eta_{jk}))]c(y_{jk},\tau_{jk}),
\label{EFppm}
\end{equation}
where the quantities $\eta_{jk}$ and $\tau_{jk}$ are, respectively, the natural and scale parameter of the distribution. The functions $a(.)$ and $c(.)$ are assumed known with $a(\eta_{jk})$ being twice differentiable with respect to $\eta_{jk}$.
\item Prior distribution for $\eta_{jk}$:
\begin{equation}
p(\eta_{jk} \mid \alpha_{jk}, \beta_{jk})= b(\alpha_{jk}, \beta_{jk})\exp [\alpha_{jk} \eta_{jk}-\beta_{jk}a(\eta_{jk})],
\label{priorippm}
\end{equation}
where $\alpha_{jk}$ and $\beta_{jk}$ with $j=1,2,\cdots,b$ and $k=1,2,\cdots,n_{j}$ are hyperparameters to be estimated. The equation above is the conjugate prior distribution of the observation equation defined in (\ref{EFppm}) and follows directly from the properties of the EF. The importance of working with conjugate priors was highlighted in the previous section. Basically, it allows us to obtain closed-form expressions for the block predictive function  without relying on  the use of intensive computational methods. The normalising constant of (\ref{priorippm}) is given by the integral,
\begin{equation*}
b(\alpha_{jk}, \beta_{jk})=\frac{1}{\int \exp [\alpha_{jk} \eta_{jk}-\beta_{jk}a(\eta_{jk})]d\eta_{jk}}.
\end{equation*}
\item Link Equation:
\begin{equation}
\lambda_{jk}=g(\eta_{jk}^{\rho})=\mathbf{F}_{jk}^{'}\bm{\theta}_{\rho}^{(j)},
\label{link}
\end{equation}
where $\mathbf{F}_{jk}$ is a known  $d \times 1$ vector of explanatory variables, $\bm{\theta}_{\rho}^{(j)}$ is a $d\times 1$ block state vector and $g(.)$ is a monotonic diffentiable link function that relates the natural parameter $\eta_{jk}$ of the observation equation to the linear predictor $\lambda_{jk}$.
\item Evolution Equation:
\begin{equation}
    \bm{\theta}_{\rho}^{(j)}=\mathbf{G}_{j}\bm{\theta}_{\rho}^{(j-1)}+\bm{\omega}_{j}\text{,}\qquad \omega_{j} \sim [0, \mathbf{W}_{j}],
    \label{sistppm}
\end{equation}
where $\mathbf{G}_{j}$ represents a $d\times d$ known evolution matrix, and $\bm{\omega}_{j}$ is a random evolution error only partially specified in term of its two first moments.
\item Prior distribution for the number of blocks $B$
\begin{equation}
p(B=b)\propto \sum_{\mathcal{C}_{1}}\prod_{j=1}^{b}c_{\rho}^{(j)},
\label{prioriB}
\end{equation}
where $\mathcal{C}_{1}$  denotes the set of all possible partitions of $I$ into exactly $b$ contiguous blocks and under the condition that $0=i_{0}<i_{1}<\cdots<i_{b}=n$.
\item Initial Information:
\begin{equation}
    (\bm{\theta}_{\rho}^{(0)} \mid D_{0}) \sim [\mathbf{m_{00}, \mathbf{C_{00}}}],
\end{equation}
where $D_{0}$ represents the initial set of information.
\end{itemize}
Equation (\ref{sistppm}) describes the Markovian evolution of the block state parameter $\bm{\theta}_{\rho}^{(j)}$. The partial specification of the errors' distribution $\omega_{j}$ only in terms of their first two moments is a standard procedure in the DGLM literature -- it allows for a straightforward inference procedure even though full distributional knowledge about the process is lost. The complete characterisation of a DGLM-PPM also requires a prior cohesion $c_{\rho}^{(j)}$ and a specification for $\mathbf{W}_{j}$. This topics will be addressed separately later on.

\subsection{Inference}
In this section we show the filtering algorithm for the DGLM-PPM. The procedure described here follows the steps of evolution and updating outlined by West, Harrison \& Migon (1985). Denote by $D_{\rho}^{(j)}$ the information set up to the $j$-th block, the inference begins by defining posterior moments of the state vector $\bm{\theta}_{\rho}^{(j-1)}$, that is:
\begin{equation*}
(\bm{\theta}_{\rho}^{(j-1)}\mid D_{\rho}^{(j-1)})\sim [\mathbf{m}_{\rho}^{(j-1)}, \mathbf{C}_{\rho}^{(j-1)}].
\end{equation*}

The moments of the block state prior, $(\bm{\theta}_{\rho}^{(j)} \mid D_{\rho}^{(j-1)})$, can be obtained directly from the Evolution Equation in (\ref{sistppm})  and are given by:
\begin{equation}
(\bm{\theta}_{\rho}^{(j)}\mid D_{\rho}^{(j-1)})\sim [\mathbf{a}_{\rho}^{(j)}, \mathbf{R}_{\rho}^{(j)}],
\end{equation}
where $\mathbf{a}_{\rho}^{(j)}=\mathbf{G}_{j}\mathbf{m}_{\rho}^{(j)}$ and  $\mathbf{R}_{\rho}^{(j)}=\mathbf{G_{j}}\mathbf{C}_{\rho}^{(j-1)}\mathbf{G_{j}}^{\prime}+\mathbf{W_{j}}$.

Now, define $D_{\rho}^{(j-1, k-1)}$ as the information set available up to the block $j-1$ and the observation $k-1$ of the $j$-th block that is being processed.
The linear predictor $\lambda_{jk}$ and the state vector $\bm{\theta}_{\rho}^{(j)}$ are assumed to have the following joint distribution specified only in terms of its first and second moments:

\begin{equation}
\left(
\begin{matrix}
\lambda_{jk}\hfill \\ \bm{\theta}_{\rho}^{(j)}
\end{matrix}
\, \middle\vert \,
D_{\rho}^{(j-1, k)}
\right) \sim \left[\left(\begin{array}{c}
f_{jk}\\
\mathbf{a_{jk}}\\
\end{array}\right),\left(\begin{array}{cc}
q_{jk} & \mathbf{F}^{\prime}_{jk}\mathbf{R}_{jk}\\
\mathbf{R}_{jk}\mathbf{F}^{\prime}_{j} & \mathbf{R}_{jk}\\
\end{array}\right)\right],
\label{lthetappm}
\end{equation}
where $ f_{jk}=\mathbf{F}^{\prime}_{j}\mathbf{R}_{jk}$ and $q_{jk}=\mathbf{F}^{\prime}_{k}\mathbf{R}_{jk}\mathbf{F}_{j}$.

Estimates of the hyperparameters $\alpha_{jk}$ and $\beta_{jk}$ of the prior distribution (\ref{priorippm}) can be obtained by matching the moments of the linear predictor, $\lambda_{jk}$, with $f_{jk}$ and $q_{jk}$ using the following relations,
\begin{equation}
f_{jk}=\mathbb{E}[\lambda_{jk}\mid D_{\rho}^{(j-1, k)}], \qquad \text{and} \qquad q_{jk}=\mathbb{VAR}[\lambda_{jk} \mid D_{\rho}^{(j-1, k)}].
\label{ft}
\end{equation}
The resulting non-linear system of equations can be solved numerically or with the help of approximations. An example of this procedure will be provided in the next section.

From Bayes's theorem and according to properties of the EF,  given the set of information $D_{\rho}^{(j-1, k)}$, the  posterior distribution of $\eta_{jk}$ will be,
\begin{equation}
p(\eta_{jk}\mid D_{\rho}^{(j-1, k-1)})=b(\alpha_{jk}+\tau_{jk}y_{jk}, \beta_{jk}+\tau_{jk})\exp[(\alpha_{jk}+\tau_{jk}y_{jk})\eta_{jk}-(\beta_{jk}+\tau_{jk})a(\eta_{jk})].
\label{posteriorippmeta}
\end{equation}


To compute the predictive distribution of the observations $\bm{Y}$ given the partition observe that, conditionally on $\rho=\{i_{0}, i_{1}, \cdots, i_{b}\}$, the joint density of the observations can be written as a product of predictive distributions, that is,

\begin{equation}
p(\bm{Y}\mid\rho)=\prod_{j=1}^{b}p(\bm{Y}_{\rho}^{(j)}),
\label{preddglmppm}
\end{equation}
where the predictive distribution $p(\bm{Y}_{\rho}^{(j)})$ associated to the observations in the $j$-th block can be calculated from
\begin{align*}
p(\bm{Y}_{\rho}^{(j)})&=\prod_{k=1}^{nj}\int p(y_{jk} \mid \eta_{jk})p(\eta_{jk})d\eta_{jk}\\
&=\prod_{k=1}^{nj}\frac{b(\alpha_{jk}, \beta_{jk})c(y_{jk}, \frac{1}{\tau_{jk}})}{b(\alpha_{jk}+\tau_{jk}Y_{jk}, \beta_{jk}+\tau_{jk})}.\numberthis
\label{prodpred}
\end{align*}



As a consequence of the product form in (\ref{prodpred}), the posterior distribution of the partition $\rho=\{i_{0}, i_{1},\cdots, i_{n}\}$ can be obtained from Equation (\ref{postrho}), which assures that the probability of any partition can be written as product of non-negative cohesions. Also, from the assumptions of the DGLM class, the observations are conditionally independent given the parameters. Thus, DGLM-PPM satisfies both conditions required for a PPM.

The updating of the linear predictor  follows directly from  (\ref{posteriorippmeta}). The posterior moments of $\lambda_{jk}$ are given by:

\[
\mathbb{E}[g(\eta_{jk})\mid D_{\rho}^{(j-1, k-1)}]=f_{jk}^{*},\qquad \text{and} \qquad \mathbb{VAR}[g(\eta_{jk})\mid D_{\rho}^{(j-1, k-1)}]=q_{jk}^{*}.
\]
In many cases the pair of equations above will not have a simple solution. In these situations, the computation of $f_{jk}^{*}$ and $q_{jk}^{*}$ may be done approximately as proposed by da-Silva, Migon \&  Correia (2011) ~\cite{dasilva2011}.

Since no assumptions are made about the distribution of the block state vector, we can not obtain a posterior distribution for $(\bm{\theta}_{\rho}^{(j)} \mid D_{\rho}^{(j-1, k)})$ without making additional hypothesis about the model.
Following the steps outlined in West \& Harrison (1997), Chapter 14, we use Linear Bayesian Estimation (LBE) to estimate the first two moments of  the posterior distribution of the block state vector. LBE is a technique that allows for the estimation of unknown non-linear functions through linear approximations. Applying the procedure to the joint distribution of $\bm{\theta}_{\rho}^{(j)}$ and $\lambda_{jk}$ we obtain:

\begin{equation*}
\hat{\mathbb{E}}[\bm{\theta}_{\rho}^{(j)} \mid \lambda_{jk},D_{\rho}^{(j-1, k)}]=\mathbf{a}_{jk}+\frac{1}{q_{jk}}\mathbf{R}_{j}\mathbf{F}_{j}(\lambda_{jk}-f_{jk}) \qquad \text{and}
\label{lbeE}
\end{equation*}
 \begin{equation*}
\hat{\mathbf{VAR}}[\bm{\theta}_{\rho}^{(j)} \mid \lambda_{jk},D_{\rho}^{(j-1, k)}]=\mathbf{R_{j}}-\frac{1}{q_{jk}}\mathbf{R_{j}}\mathbf{F_{j}}\mathbf{F_{j}^{\prime}}\mathbf{R_{j}}.
\label{lbeV}
\end{equation*}
Then, using the Law of Iterated Expectation, we have that $(\bm{\theta}_{\rho}^{(j)} \mid D_{\rho}^{(j-1, k)})\sim [\mathbf{m}_{jk}, \mathbf{C}_{jk}]$, where
\begin{equation}
\mathbf{m}_{jk}=\mathbf{a}_{jk}+\frac{1}{q_{jk}}\mathbf{R}_{jk}\mathbf{F}_{jk}(f_{jk}^{*}-f_{jk})\numberthis \label{mtppm} \qquad
\end{equation}
and
\begin{equation}
\mathbf{C}_{jk}= \mathbf{R}_{jk}-\frac{1}{q_{jk}}\left[\mathbf{R}_{jk}\mathbf{F}_{jk}\mathbf{F}_{jk}^{\prime}\mathbf{R}_{jk}\left(1-\frac{q_{jk}^{*}}{q_{jk}}\right)\right].
\label{Cppm}
\end{equation}

The inference procedure goes as follows: within each block, there is no parametric evolution. The information set available at the beginning of the block $[i_{j-1}i_{j}]$ is denoted by $D_{\rho}^{(j-1, 0)}$ and the updating procedure is started by taking $\mathbf{a}_{j1}=\mathbf{m}_{(j-1,n_{j-1})}$ and $\mathbf{R}_{j1}=\mathbf{C}_{(j-1,n_{j-1})}$. After each observation is processed, the values of $\mathbf{a}_{jk}$ and $\mathbf{R}_{jk}$ are updated to $\mathbf{a}_{jk}=\mathbf{m}_{(j, k-1)}$ and $\mathbf{R}_{jk}=\mathbf{C}_{(j, k-1)}$. Once every observation in the $j$-th block is handled, the evolution equation is applied to the state vector, and the inference proceeds to the next block. This cycle is carried out until all the observations of every block are processed. Naturally, the complete specification of a DGLM-PPM requires an initial state denoted by:


\begin{equation}
    (\bm{\theta}_{\rho}^{(0)} \mid D_{0}) \sim [\mathbf{m_{00}, \mathbf{C_{00}}}],
\end{equation}
where $D_{0}$ represents the initial set of information.

Table (\ref{outrasdists}), below, displays the parameters $\alpha_{jk}$, $\beta_{jk}$, $f^{*}_{jk}$ and $q^{*}_{jk}$ for some of the distributions within the EF. In Section 4, we will work out the Poisson DGLM-PPM as an illustration.

\begin{table}[H]
\begin{tabular}{ccccc}
\hline
\textbf{Distribution}                  & $\alpha_{jk}$                      & $\beta_{jk}$                                             & $f_{jk}^{*}$                                                                           & $q_{jk}^{*}$   \\                                                                    \\ \hline
Poisson ($\mu_{jk}$)                   & $\frac{1}{q_{jk}}$                 & $\frac{\exp(-f_{jk})}{q_{jk}}$                           & $\log\left(\frac{\alpha_{jk}+y_{jk}}{\beta_{jk}+1}\right)$                             & $\frac{1}{y_{jk}+\alpha_{jk}}$                                                     \\ \\
Normal ($\mu_{jk}, V$)                 & $f_{t}$                            & $q_{t}$                                                  & $\frac{q_{jk}y_{jk}+\alpha_{jk}V}{V+\beta_{jk}}$                                       & $\frac{\beta_{jk}V}{V+\beta_{jk}}$   \\                                              \\
Binomial ($k_{jk}, \mu_{jk}$)          & $\frac{+\exp(f_{jk})}{q_{jk}}$     & $\frac{1+\exp(-f_{jk})}{q_{jk}}$                         & $\log\left(\frac{\alpha_{jk}+y_{jk}}{\beta_{jk}+k_{jk}-y_{jk}}\right)$                  & $\frac{1}{y_{jk}+\alpha_{jk}}+\frac{1}{\beta_{jk}+k_{jk}-y_{jk}}$                  \\ \\
Neg. Binom. ($\pi_{jk}, \lambda_{jk}$) & $\frac{1-\exp(-f_{jk})}{q_{jk}}$   & $\frac{1-2\exp(f_{jk})+\exp(2f_{jk}}{\exp(f_{jk}q_{jk}}$ & $\log\left(\frac{\beta_{jk}+y_{jk}}{\beta_{jk}+\lambda_{jk}+y_{jk}+\alpha_{jk}}\right)$ & $\frac{1}{y_{jk}+\beta_{jk}}+\frac{1}{\beta_{jk}+\lambda_{jk}+y_{jk}+\alpha_{jk}}$ \\ \\
Gamma($r_{jk}, s_{jk}$)            & $\frac{-\alpha_{jk}}{\beta_{jk}}$ & $\frac{\alpha_{jk}^{2}}{\beta_{jk}}$                    & $\frac{-\alpha_{jk}}{\beta_{jk}+y_{jk}}$                                               & $\frac{\alpha_{jk}}{(\beta_{jk}+y_{jk})^{2}}$                                      \\ \hline
\end{tabular}
\\
\caption{Approximate values for the parameters of the DGLM-PPM}
\label{outrasdists}
\end{table}

\subsection{The prior cohesion}
Some classes of clustering models require the specification of a prior cohesion function.  In the absence of information, the modeler may choose  non-informative discrete cohesions such as an uniform prior. For exchangeable partition models, the Dirichlet and Pitman-Yor process are popular choices (for an in-depth discussion, see Pagananin (2020) ~\cite{paganin2020}). In the case of PPM models that only allow for contiguous blocks, truncated geometric priors are a common pick -- the most famous one being Yao's prior ~\cite{yao84} .   In this work we chose to work with one of such priors. Let $\pi$ define the change-point probability at any given instant $t$, we define the prior cohesion for block $j$ as
\begin{equation}
 c_{\rho}^{(j)}= \begin{cases}
     \pi(1-\pi)^{i-j-1}, \qquad \text{if}\qquad j<n\\
      (1-\pi)^{i-j-1}, \qquad \text{ } \text{ if}\qquad j=n,\\
   \end{cases}
   \label{yao}
\end{equation}
for all $i,j \in \rho$ such that $i<j$. Equation (\ref{yao}) defines a discrete renewal process wherein the change-points are represented by an \textit{iid} Bernoulli sequence and are independent of one another. That is, by assuming Yao's prior, one considers that past realisations of the process do not convey information about its future.  Under this setting, it can be shown that the probability associated to any partition is given by,
\begin{equation*}
    p(\rho=\{i_{1}, \cdots, i_{b}\})=\pi^{b-1}(1-\pi)^{n-b},
\end{equation*}

Another direct consequence of Equation (\ref{yao}) is that the prior distribution of the random variable $B$ denoting the number of blocks in the partition follows a binomial distribution, that is
\begin{equation}
    P(B=b)=\binom{n-1}{b-1}\pi^{b-1}(1-\pi)^{n-b}.
\end{equation}
Loschi et al. (2002) ~\cite{loschi2003} suggest to assign a Beta prior distribution with parameters r and s, $\mathcal{B}(r, s)$,  for the probability $\pi$. This allows for exact calculation of the relevant quantities associated to Yao's prior. Suppose $\pi\sim \mathcal{B}(r, s)$ then the random variable  $B-1$  follows a Beta Binomial distribution with parameters $n-1$, $r$ and $s$, which, in turn,  imply a prior mean for the number of change-points  given by
\begin{equation}
    \mathbb{E}(B-1)=(n-1)\frac{r}{r+s}.
    \label{priormeanB}
\end{equation}
The marginal posterior distribution of $\rho$ given the data can be written as
\begin{equation*}
    p_{\pi}(\rho \mid \bm{Y})= \int_{\Pi} p(\rho \mid \bm{Y}) p(\pi)d\pi,
\end{equation*}
where $p(\rho \mid \bm{Y})$ can be calculated from  (\ref{postrho}). Therefore, the posterior distribution of $\rho$ can be expressed as
\begin{equation}
    p_{\pi}(\rho \mid \bm{Y}) \propto \left[\prod_{j=1}^{b}p_{j}(\bm{Y}_{\rho}^{(j)})\right]\mathcal{B}(b+r-1, n+s-b),
    \label{postrhomarg}
\end{equation}

From the results above, it is possible to obtain the exact posterior relevances involved in the calculation of  (\ref{esperancapost}). Nevertheless, as already discussed in Section 2,  the computation of the products and sums requires a high computational effort. An alternative approach for this problem is the Gibbs sampling scheme that will be introduced next.

\subsection{Gibbs Sampling}
In this section, we present the Gibbs sampling algorithm proposed by Barry \& Hartigan (1993) ~\cite{barry93} and used by Loschi \& Cruz (2002) \cite{loschi2002} to carry out inferences on the partition and  parameters of a PPM.

 Let $\mathbf{U}$ be an auxiliary  random vector  of length $n-1$ whose $i$-th component is defined as:

$$
U_{i}=
\begin{cases}
1, \qquad \text{ if} \qquad \theta_{i}=\theta_{i+1}\\
0, \qquad \text{ if} \qquad \theta_{i}\neq \theta_{i+1}
\end{cases}
$$

Each element $U_{i} \in U$ for $i=1,2,\cdots,n-1$ is an indicator variable that takes on zero or one whether or not  the process goes through a change-point at the  $i$-th observation. Observe that any  partition  $\rho=\{i_{0}, i_{1}, \cdots, i_{b}\}$ is  completely defined by the vector $U=(U_{1},\cdots,U_{n-1})$.
 The Gibbs Sampler begins with  $\mathbf{U}^{0}=(U_{1}^{0},\cdots,U^{0}_{n-1})$. For every step $s>1$,
a new vector  $U^{s}=(U_{1}^{s},\cdots,U^{s}_{n-1})$ is generated such that the value of each element  $U_{r}^{s}$ is determined conditionally on the values of the other variables according to the following density:
\[
f(U_{r} \mid U_{1}^{s},\cdots, U_{r-1}^{s}, U_{r+1}^{s-1},\cdots, U_{n-1}^{s-1}; \bm{Y}),
\]
with $r=1,\cdots,n-1$.

To generate new samples from the partition $\rho$, Loschi \& Cruz (2002) propose using the ratio below,

\begin{equation}
R_{r}=\frac{P(U_{r}=1 \mid A_{r}^{s}, \pi, \bm{Y})}{P(U_{r}=0 \mid A_{r}^{s},\pi, \bm{Y})},
\label{rgibbs}
\end{equation}
where $A_{r}^{s}=\{U_{1}^{s}=u_{1},\cdots, U_{r-1}^{s}=u_{r-1},U_{r+1}^{s-1}=u_{r+1},\cdots, U_{n-1}^{s-1}=u_{n-1}\}$ and $r=1,\cdots,n-1$. Let $b$ be the number of blocks when $U_{r}=0$ and assume Yao's prior cohesion defined in (\ref{yao}) with a $\mathcal{B}(r,s)$ prior for the parameter $\pi$. Then, from  (\ref{postrhomarg}), Equation (\ref{rgibbs}) becomes:

\begin{equation}
    R_{r}=\frac{P(\mathbf{Y}\mid \rho, r, s)}{P(\mathbf{Y}\mid \rho^{\prime}, r, s)}=\frac{\prod_{j}^{b^{\prime}}p_{j}(\bm{Y}_{\rho^{\prime}}^{(j)})\text{Beta}(b+r-2, n+s-b+1)}{\prod_{j}^{b}p_{j}(\bm{Y}_{\rho}^{(j)})\text{Beta}(b+r-1, n+s-b)}
    \label{gibbsratio}
    \end{equation}
where $\text{Beta}(.)$ denotes the Beta function, $\rho=\{U_{1}=u_{1}, \cdots, U_{r}=0, \cdots, U_{n-1}=u_{n-1}\}$, $\rho^{\prime}=\{U_{1}=u_{1}, \cdots, U_{r}=1, \cdots, U_{n-1}=u_{n-1}\}$ and $b^{\prime}=b-1$. Rewriting the Beta function in terms of Gamma functions and using the recurrence relation $\Gamma(x+1)=x\Gamma(x)$, Equation (\ref{gibbsratio}) simplifies to,
\begin{equation}
    R_{r}=\frac{\prod_{j}^{b^{\prime}}p_{j}(\bm{Y}_{\rho^{\prime}}^{(j)})(n+s-b)}{\prod_{j}^{b}p_{j}(\bm{Y}_{\rho}^{(j)})(b+r-2)}.
    \end{equation}

A criterion for accepting or rejecting the values obtained for $U_{r}^{s}$  is
$$
U_{r}^{s}=
\begin{cases}
1 \text{, if} \qquad R_{r}\geq \frac{1-u}{u}\\,
0 \text{, otherwise}
\end{cases}
$$
where the values of $u$ are drawn from a standard Uniform distribution.

Let $M$ be the size of the MCMC chain generated (that is, the number of vectors $U$) and $M^{\prime}$ the number of vectors containing the block $j$. The posterior relevance $r^{*}(j)$ is defined as
\begin{equation}
    r^{*}(j)=\frac{M^{\prime}}{M}.
\end{equation}
The product estimates associated to the parameters of interest can be obtained from  (\ref{esperancapost}) using the relevances computed according to the equation above. The number of blocks $B$ in the partition can be obtained from




\begin{equation}
    B=1+n-\sum_{r=1}^{n-1}U_{r},
    \label{nblocexp}
\end{equation}
and the posterior change-point probability is given by
\begin{equation}
    \displaystyle Pr(y_{r} \text{ is a change point} \mid \mathbf{Y}, \mathbb{\varrho})=\frac{\sum_{i=1}^{M}U_{r}^{(i)}}{M},
    \label{changprobexp}
\end{equation}
where $\varrho=\{\rho^{(1)}, \rho^{(2)},\cdots, \rho^{(i)}, \cdots, \rho^{(M)}\}$ and $U_{r}^{(i)}$ corresponds to  the $r$-th observation of the $i$-th partition, $\rho^({i})$, sampled by the Gibbs algorithm.
\subsection{Discount Factor}

In a DGLM setting, the usual strategy to specify the variance matrix of the evolution equation is via discount factors (denoted, here, by $\delta$). This approach is very appealing since it handles the uncertainty about the state parameters elegantly and straightforwardly. A decrease in the discount factor is associated with a larger variance $\mathbf{W}_{j}$ in the state evolution and, consequently, with more  uncertainty in the parameters. In most cases, the specification of $\delta$ is carried on through a sensitivity analysis in which the chosen values of the discount factor are close to one. This is a reasonable procedure since it is expected that a time series should retain most of its information when transitioning from one observation to the next.
 That is not the case for the DGLM-PPM, however. Since we now have a clustering structure that aggregates several observations and the evolution equation is only applied between different blocks, higher values of $\delta$ would imply very little uncertainty about the future state of the process. Therefore, the block structure is likely to induce smaller values of the discount factor for the DGLM-PPM in relation to the conventional DGLM.

 Although a sensitivity analysis taking into account different values of $\delta$ is possible for the DGLM-PPM, in practice it would demand a high computational effort. In order to circumvent this issue, we propose to append a Metropolis step within the Gibbs Sampling scheme described in the last section. This way we can obtain point and interval estimates of the discount factor in a reasonable computer time. The posterior distribution for $\delta$ is given by the following expression:

\begin{equation}
    p(\delta \mid D^{(b)}) \propto \left[\prod_{j=1}^{b} \prod_{k=1}^{n_{j}}p(Y_{jk} \mid D^{(j-1, k-1)}, \rho ,\delta)\right]p(\delta),
    \label{postdelta}
\end{equation}

 where $p(\delta)$ represents a prior distribution for the discount factor and $D^{(b)}$ is the complete information set up to the last block $b$. Since $\delta \in [0,1]$, the Beta distribution arises as natural candidate to be the prior distribution for the discount factor $\delta$ . Samples from  (\ref{postdelta}) can be obtained efficiently using the well known Adaptative Rejection Metrolis Sampling (ARMS) ~\cite{gilks95} algorithm.

\section{Poisson DGLM-PPM}

Consider a time series of counts $Y_{jk}$ with $j=1, 2, \cdots, b$ and $k=1, 2, \cdots, n_{j}$ such that $Y_{jk} \mid \mu_{jk} \sim \text{Poisson}(\mu_{jk})$ and $\mu_{jk}>0$. The Poisson distribution belongs to the EF with $\tau=1$, $a(\eta_{t})=\exp(\eta_{t})$, $c(y_{t}, \tau)=\frac{1}{y_{t}!}$ and $\eta_{t}=\log(\mu_{t})$. From (\ref{link}) and properties of the EF, the canonical link function is given by $\lambda_{jk}=g(\eta_{jk})=F_{jk}^{\prime}\bm{\theta}_{jk}$. Thus, assuming $g(\eta_{jk})=\eta_{j}$, we obtain,
\begin{equation*}
\log(\mu_{jk})=F_{jk}^{\prime}\bm{\theta}_{jk}
\end{equation*}

Equation (\ref{priorippm}) implies a $\text{Gamma}(\alpha_{jk}, \beta_{jk})$ conjugate prior distribution for $\mu_{jk}$, that is
\begin{equation*}
p(\mu_{jk} \mid D_{t-1})=\frac{\beta_{jk}^{\alpha_{jk}}}{\Gamma(\alpha_{jk})}\mu_{jk}^{\alpha_{jk}-1}exp(-\beta_{jk}\mu_{jk}).
\end{equation*}
The hyperparameters $\alpha_{t}$ and $\beta_{t}$ can be elicited in terms of the pair ($f_{t}$, $q_{t}$)  using the following relations:
\begin{equation*}
f_{jk}=\mathbb{E}[\log(\mu_{jk} \mid D_{j-1, k})], \qquad \text{and} \qquad q_{jk}=\mathbb{VAR}[\log(\mu_{jk} \mid D_{j-1, k})].
\end{equation*}

Thus, $f_{jk}=\psi(\alpha_{jk})-\log(\beta_{jk})$ and $q_{jk}=\psi^{\prime}(\alpha_{jk})$, where $\psi$ and $\psi^{\prime}$ are the Gamma and Digamma functions, respectively. West \& Harrison (1997) suggest to use the following first order approximations for the digamma and trigamma  functions: $\psi(x)=\log(x)$ and $\psi^{\prime}(x)=\frac{1}{x}$. Solving the resulting system of equations we obtain $\alpha_{jk}=\frac{1}{q_{jk}}$ and $\beta_{jk}=\frac{\exp(-f_{jk})}{q_{jk}}$.

From the conjugacy assumption, the posterior distribution of $\mu_{jk}$ given the partition will be
\begin{equation*}
    p(\mu_{jk}\mid D_{jk}, \rho) \sim \text{Ga}(\alpha_{jk}+y_{jk}, \beta_{jk}+1).
\end{equation*}
The equation above implies that the posterior moments of the linear predictor can be approximated by the following expressions,
\begin{equation*}
    f^{*}_{jk} \approx \log\left(\frac{y_{jk}+\alpha_{jk}}{\beta_{jk}+1}\right)\qquad \text{and} \qquad q^{*}_{jk}=\frac{1}{y_{jk}+\alpha_{jk}}
\end{equation*}
From  (\ref{prodpred}) the block predictive distribution takes the form:
\begin{equation*}
    p_{j}(\bm{Y}_{\rho}^{(j)})=\prod_{i=1}^{n_{j}}\frac{\Gamma(\alpha_{jk}+y_{jk})}{y_{jk}!\Gamma(\alpha_{jk})}\frac{\beta_{jk}^{\alpha_{jk}}}{(1+\beta_{jk})^{y_{jk}+\alpha_{jk}}}
\end{equation*}

Thus, assuming Yao's prior cohesion as defined in Section (3.3), the posterior distribution of the partition $\rho$ given the observations $\bm{Y}$ can be expressed as
\begin{equation*}
    p_{\pi}(\rho\mid \bm{Y}, \delta) \propto \left[\prod_{j=1}^{b}\prod_{i=1}^{n_{j}}\frac{\Gamma(\alpha_{jk}+y_{jk})}{y_{jk}!\Gamma(\alpha_{jk})}\frac{\beta_{jk}^{\alpha_{jk}}}{(1+\beta_{jk})^{y_{jk}+\alpha_{jk}}}\right]\text{Beta}(b+r-1, n+s-b).
\end{equation*}
In the same way, considering a $\mathcal{B}(r,s)$ prior, the posterior distribution of the discount factor $\delta$ given the data and the partition will be given by
\begin{equation*}
    p(\delta \mid \bm{Y}, \rho) \propto \left[\prod_{j=1}^{b}\prod_{i=1}^{n_{j}}\frac{\Gamma(\alpha_{jk}+y_{jk})}{y_{jk}!\Gamma(\alpha_{jk})}\frac{\beta_{jk}^{\alpha_{jk}}}{(1+\beta_{jk})^{y_{jk}+\alpha_{jk}}}\right](1-\delta)^{r-1}\delta^{s-1}.
\end{equation*}

Finally, updating of the state vector can be done via equations (\ref{mtppm}) and (\ref{Cppm}).
\section{Simulation Study}

A simulation study was carried out to evaluate the properties of the DGLM-PPM  and compare it to the conventional DGLM. We generated $L=500$ samples of size $n=100$ observations from a Poisson Local Level Model (PLLM). The equations describing the generating process are given below,
\begin{align}
    y_{t}&\sim \text{Poisson}(\exp(\gamma_{t})) \label{obsPPML}\\
    \gamma_{t}&=\phi \gamma_{t-1}+\eta_{t}, \qquad \eta_{t}\sim N[0, W_{t}] \numberthis,
    \label{PPML}
\end{align}
where $\gamma_{t}$ represents the level of the process at time $t$, $\phi$ is a constant and $\eta_{t}$ a normally distributed stochastic error with variance $W_{t}$. To ensure the generation of stable stationary samples, we  set $\phi=0.99$ and $W_{t}=0.001$. Since the objective is to analyse time series with change-points, jumps were introduced in the level $\gamma_{t}$ of the processes at the instants $t=20$, $t=40$, $t=60$ and $t=80$, totalling four change-points (or five blocks).  These jumps were created so that the expected value of the time series shifts by 100 units at each jump.

To model the PPML processes with jumps defined by (\ref{obsPPML}) and (\ref{PPML}), we used the Poisson DGLM-PPM described in Section 4 specified according to the following configuration: $G_{j}=1$, $F_{j}=0.99$ and  $\bm{\theta}_{\rho}^{(j)}=\gamma_{\rho}^{(j)}$ for all $j$. No prior knowledge was assumed about the discount factor, thus we considered a non-informative  $\mathcal{B}(1.1)$ prior for $\delta$. Finally, the initial state was partially specified in terms of the first two moments as following,  $(\bm{\theta}_{\rho}^{0} \mid D_{0}) \sim (1, 10)$.



In regard to the prior cohesion, we assume Yao's prior with a non-informative $\mathcal{B} (1,1)$ prior for the index parameter $\pi$. From Equation (\ref{priormeanB}), under this configuration, the average prior expected number of blocks in the partitions is 49.5.  So, with this choice of parameters we should see the average posterior number of blocks shifts away from the prior's specification towards smaller values. The  Gibbs Sampling scheme described in previous sections was used to sample from the partition $\mathbf{U}$ and $\delta$. As each replication of the Monte Carlo experiment is very demanding from the computational point of view, we looked for a minimum viable configuration for the procedure. Preliminary analysis using the tools available on the R \texttt{coda} package \cite{coda} showed fast convergence and low correlation for the MCMC chains (Loschi \& Cruz (2002) \cite{loschi2002} report a similar result in an akin setting).
 Thus, we opted for generating single chains of 4000 samples with a burn-in period of 1000 iterations and a lag of 3 for a net sample size of $M=1000$\footnote{The computing time for each Monte Carlo replication under the setting described is around 15 minutes using an Intel i7-4790k 4GHZ processor with 16Gb of Ram memory}.

\begin{figure}[H]
\centering
\includegraphics[height=10cm, width=12cm]{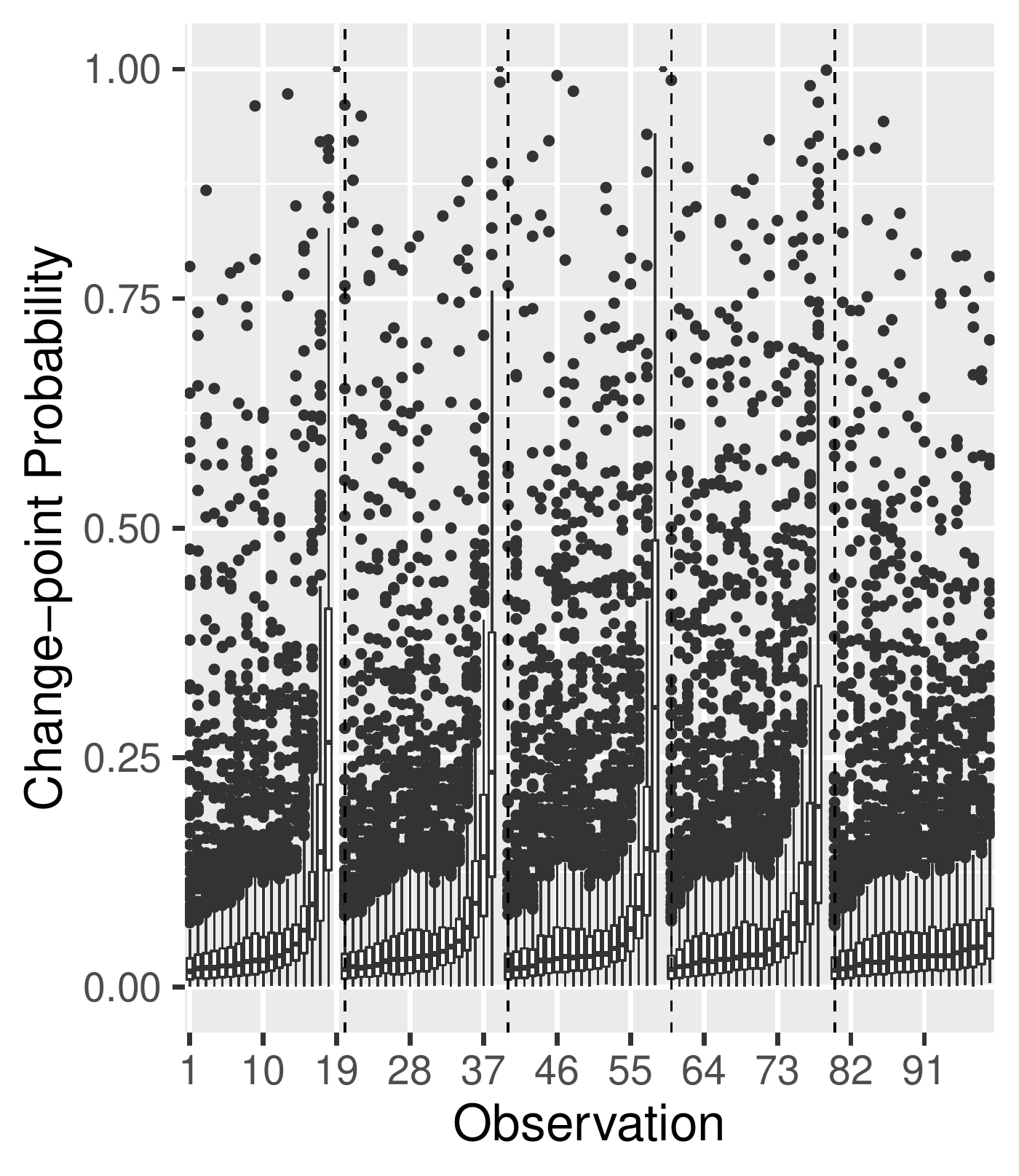}
\caption{Box Plots of the change-point probabilities associated to each observation. The vertical dashed lines indicate the change-points introduced on the generating process. }
\label{dglmsimprobchange}
\end{figure}

Point estimates of the probability that, at any given observation $y_{r}$, $r \in (1,2,\cdots, 100)$, of the time series there is a change-point can be computed according to Equation (\ref{changprobexp}). The 500 values obtained for each observation  are organised in box-plots and displayed in Figure (\ref{dglmsimprobchange}). The results show good performance by the proposed model: as expected, the change-point probabilities are equal to one across all 500 Monte Carlo replications at the jumps artificially inserted in the process. For all the remaining points, the DGLM-PPM assigns small probabilities (around 8\% on average).

Two important aspects of the DGLM-PPM are the estimation of the discount factor, $\delta$, and the posterior number of blocks in the partition, $B$. Point estimates of $\delta$ were calculated straightforwardly using the samples drawn from the  posterior distribution (\ref{postdelta}) using the ARMS algorithm; whereas an estimation of $B$ was derived from (\ref{nblocexp}) according to the procedure outlined in the previous sections. Table (\ref{deltasim}) summarises the results of the experiment. For both $\delta$ and $B$ we report the posterior mean  along with 95\% Highest Posterior Density (HPD) Credible Intervals (CI).  In regard to the discount factor, we also present the Average Standard Error (ASE) and the Standard Deviation of the punctual estimates (SDE). As the number of Monte Carlo replications increases, we expect these two quantities to come closer. All the values displayed, including the HPD interval limits, are sample averages calculated from the 500 replications used in this experiment.  For comparison purposes, we also present the estimates of the discount factor obtained for the conventional DGLM.

\begin{table}[H]
\centering
\begin{tabular}{cccccc}
\hline
\multicolumn{1}{l}{} & \textbf{\begin{tabular}[c]{@{}c@{}}Mean $\delta$\\ (95\% CI)\end{tabular}} & \textbf{ASE $\delta$} & \textbf{SDE $\delta$} & \textbf{\begin{tabular}[c]{@{}c@{}}Mean Posterior Number \\ of blocks\\ (95\% CI)\end{tabular}} & \textbf{\begin{tabular}[c]{@{}c@{}}Median Posterior \\ Number of Blocks\end{tabular}} \\ \hline
\textbf{DGLM-PPM}    & \begin{tabular}[c]{@{}c@{}}0.009\\ (0.0004, 0.023)\end{tabular}               & 0.007                    & 0.005                 & \begin{tabular}[c]{@{}c@{}}15.14\\ (6.75, 17.85)\end{tabular}                                  & 15                                                                                   \\
\textbf{DGLM}        & \begin{tabular}[c]{@{}c@{}} 0.16\\ (0.12,  0.20)\end{tabular}                & 0.021                     & 0.023                 & -                                                                                               & -                                                                                     \\ \hline
\end{tabular}
\caption{Point and interval estimates for the discount factors and the posterior number of blocks in the partition}
\label{deltasim}
\end{table}



\begin{figure}[H]
    \centering
    \includegraphics[width=12cm, height=8cm]{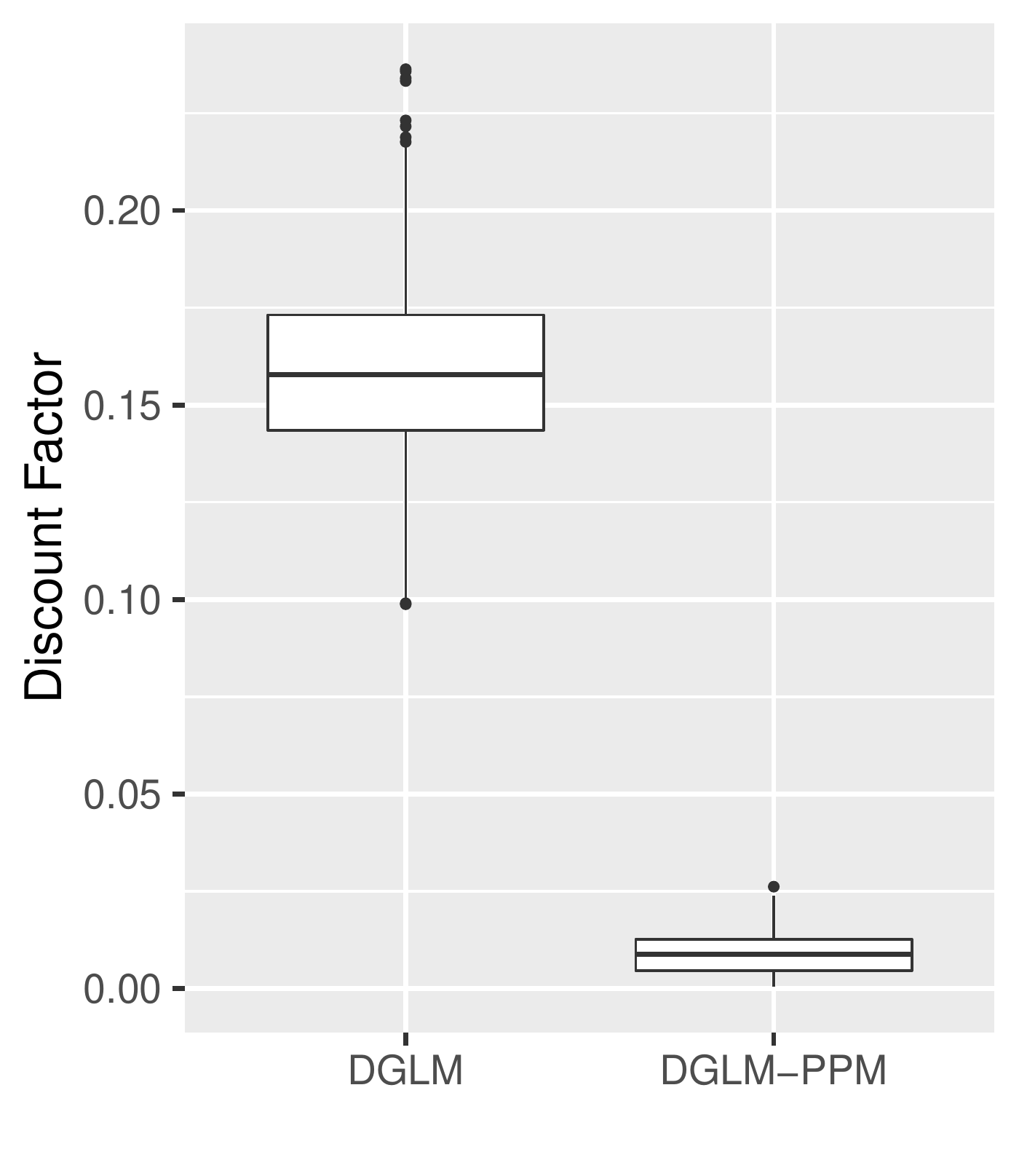}
    \caption{Comparison of the discount factors estimated via ARMS. On the left panel, estimates for the conventional DGLM, and on the right, estimates for the DGLM via PPM}
    \label{discfbp}
\end{figure}

The results show a higher discount factor for the conventional DGLM in relation to the DGLM-PPM -- an outcome that was expected and is in agreement with the discussion carried on earlier in this text. We also observe that the values of ASE and SDE are close for both the DGLM-PPM and Conventional DGLM estimates of $\delta$. The empirical distributions of the discount factor estimates can be visualised in the box-plots of Figure (\ref{discfbp}).


\begin{figure}[H]
    \centering
    \includegraphics[width=12cm, height=8cm]{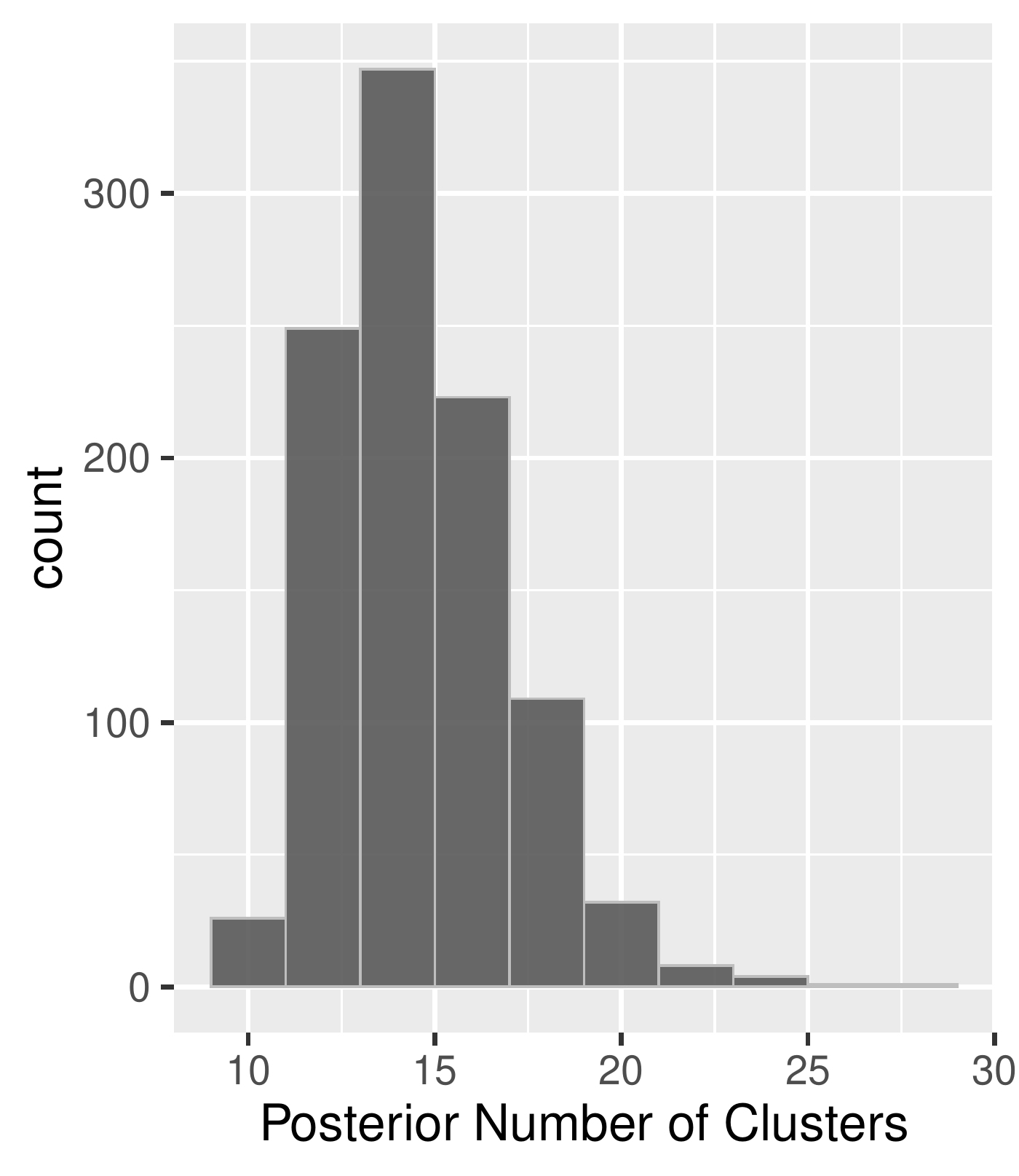}
    \caption{Empirical distribution of the posterior expected number of blocks in the partition.}
    \label{nbloc}
\end{figure}

 Figure (\ref{nbloc}) shows the empirical distribution of $B$. We find that the mean posterior number of blocks is approximately 15 with lower and upper bounds of the HPD 95\% CI around 6 and 17 blocks. This result is not completely satisfactory from an inferential point of view as we would expect to estimate  only five clusters. A possible explanation for this discrepancy is that the generating process is very close to non-stationarity and, therefore, has a lot of variability, which can conduct to the detection of undesired change-points. It should be noted, however,  that we observe a significant shift from the prior expected number of blocks, thus suggesting that the proposed model can, effectively, move towards the real $B$ even if the prior cohesion chosen misrepresents the data in consideration.


To assess the degree of similarity between the parameters estimated by the DGLM-PPM and the real underlying process we proceeded to compare the estimates obtained for the posterior mean of the state vector to the real ones generated for this study. Remember that the state vector is only partially specified in the terms of its first two moments, that is: $\theta_{jk}\sim [\mathbf{m}_{jk}, \mathbf{C}_{jk}]$. Our interest here lies on the posterior mean $\mathbf{m}_{jk}$. Denote the estimates of $\mathbf{m}_{jk}$ by $\mathbf{\hat{m}}_{jk}$, the Relative Bias (RB) associated to this estimator, in percent, will be given by:
\begin{equation}
RB(\%)=\frac{100(\mathbf{\hat{m_{jk}}}-\mathbf{m}_{jk})}{\abs{\mathbf{m}_{jk}}}
\end{equation}

\begin{figure}[ht]
    \centering
    \includegraphics[width=15cm, height=8cm]{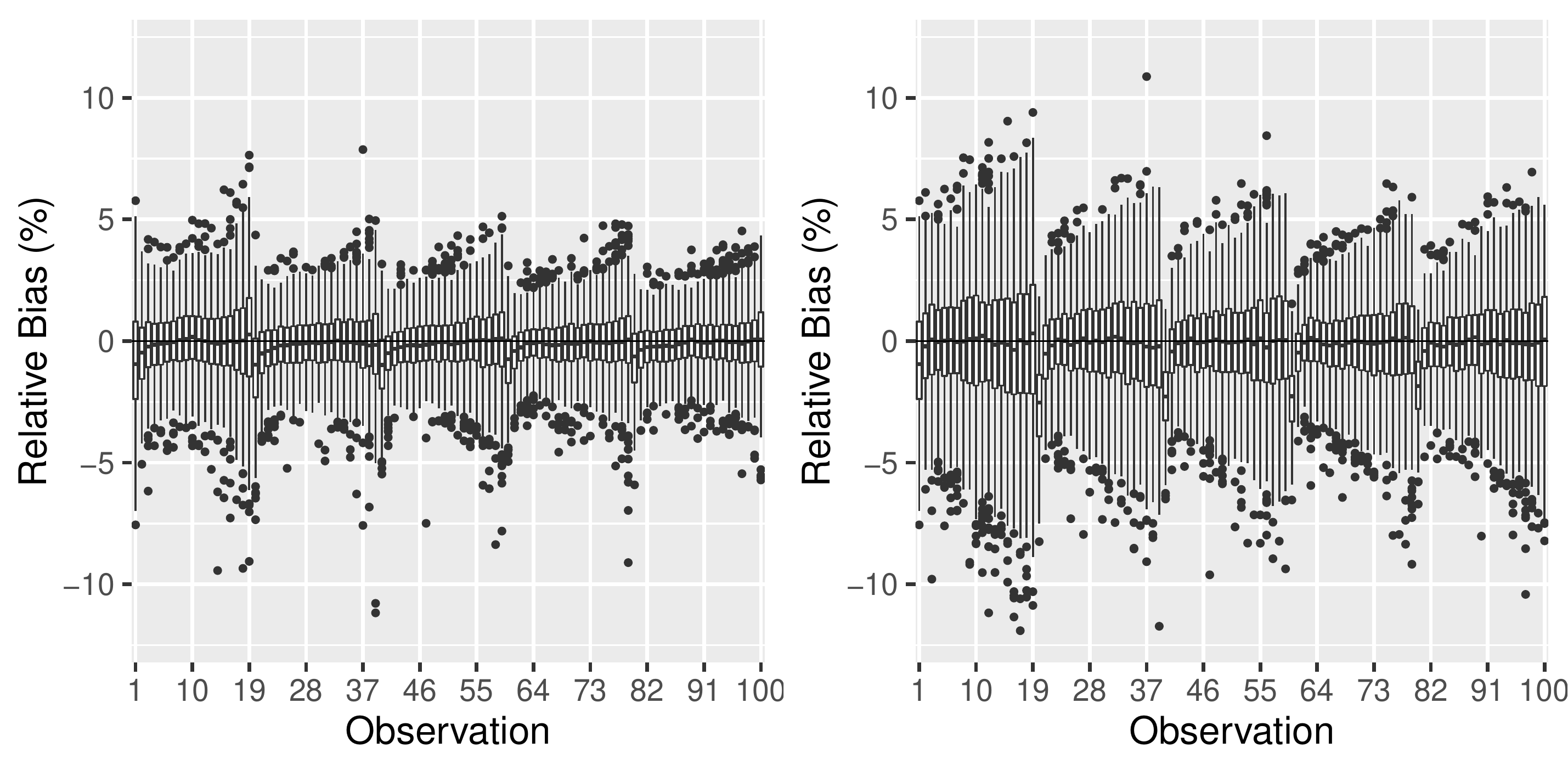}
    \caption{Box Plots of the Relative Bias associated to the estimation of the mean of the posterior state vector for the DGLM-PPM (left-panel) and the DGLM (right-panel). The horizontal black line is set at $RB=0$}
    \label{rbias}
\end{figure}

We computed the RB in the estimates of $\mathbf{m}_{jk}$ for each observation over  all Monte Carlo samples. For comparison, the procedure was also performed with the conventional DGLM using the same simulated data. The results obtained were organised in box plots and are  displayed in the Figure (\ref{rbias}). The left and right panels reports the RB associated to the DGLM-PPM and the conventional DGLM, respectively.

\begin{figure}[ht]
    \centering
    \includegraphics[width=15cm, height=8cm]{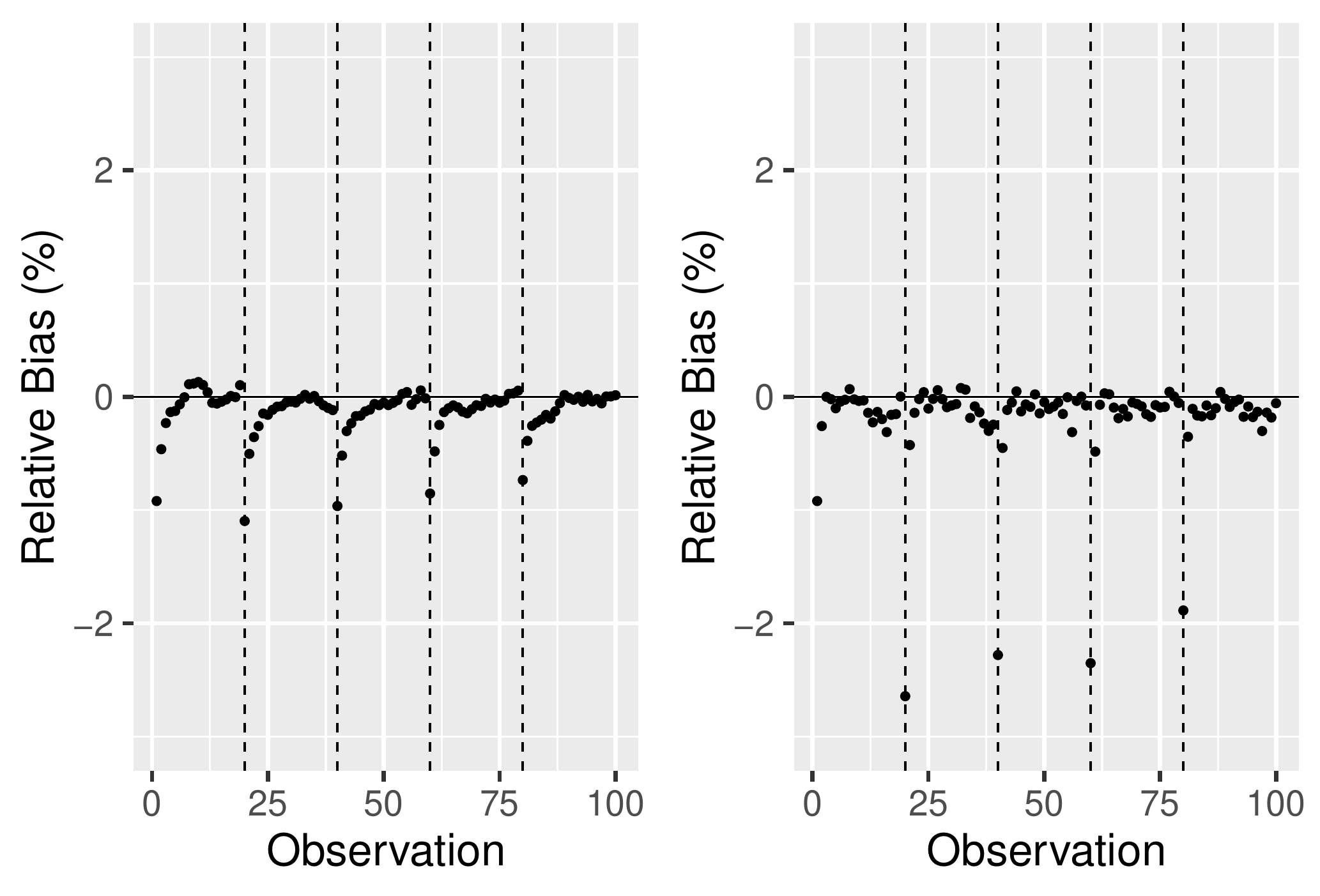}
    \caption{Mean RB for the estimates of the posterior mean of the state vector for the DGLM-PPM and the DGLM. The dashed lines indicate $RB=0$.}
    \label{rbiasmean}
\end{figure}

Overall, taking both models into account, the biases are  small. We find a mean RB of -0.13\% for the DGLM-PPM and of -0.20\% for the conventional DGLM, giving a slight advantage to our model. Considering the median RB we obtain -0.061\%  and -0.09\% respectively. Thus, this experiment shows that the DGLM-PPM provides an adjustment  at least as good as the traditional DGLM.  Larger biases are observed for the first observations after a change-point, which is an expected outcome since the process may take some time to adapt to a regime switch. An important question that arises is if this adaptive process is more efficient using the DGLM-PPM. The answer is yes! To  explore this issue, we plot  only the mean RB ($\overline{RB_{r}}$)\footnote{$\overline{RB_{r}}=\frac{1}{500}\sum_{i=1}^{500}RB_{r_{i}}$} associated to each observation $r$ (Figure \ref{rbiasmean}). Again, results obtained for the DGLM-PPM are on the left and for the conventional DGLM on the right one. The panels show a small edge in favour of the DGLM-PPM -- as can be observed by the RB values, the estimates for $\mathbf{m}_{jk}$ provided by the proposed model are markedly better  than those of the conventional DGLM for the points immediatly after the jumps. This outcome is probably due to the lower discount factor induced by the block structure that allows the DGLM-PPM to better adapt to the regime switches.

\begin{figure}[H]
    \centering
    \includegraphics[width=15cm, height=8cm]{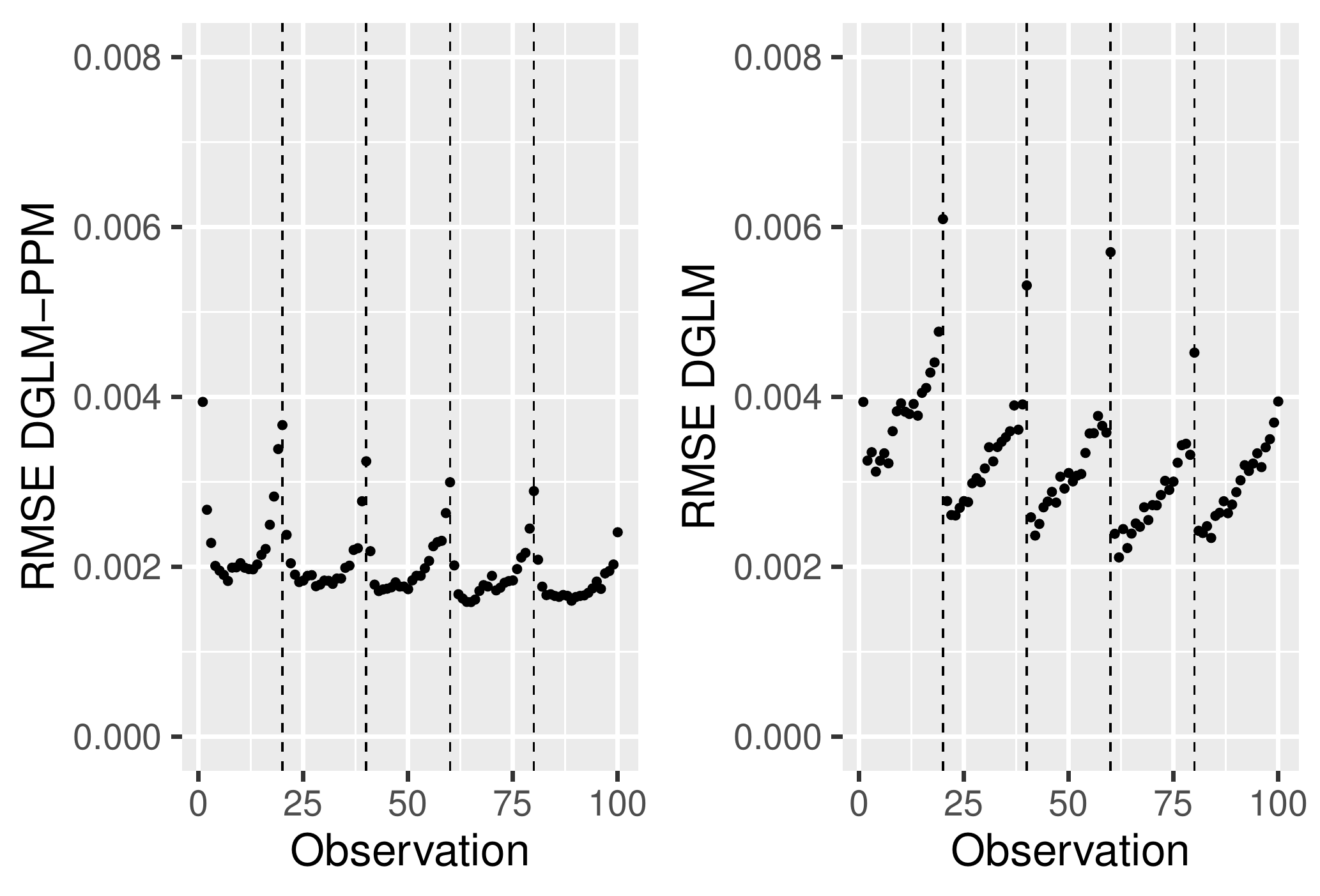}
    \caption{RMSE calculated for the estimates obtained with the DGLM-PPM (left) and DGLM (right). The vertical lines indicate the change-points}
    \label{rmse}
\end{figure}

Finally, we want to explore the variability of the estimations. Figure (\ref{rbias}) suggests that the estimates of $\mathbf{m}_{jk}$ present higher variability in the case of the conventional DGLM when compared to the DGLM-PPM.
 To  further illustrate this feature, we display in Figure (\ref{rmse}), the Root of the Mean Squared Error (RMSE) \footnote{$RMSE=\sqrt{\frac{\sum_{i=1}^{500}(\mathbf{\hat{m}}_{jk}-\mathbf{m}_{jk})^{2}}{500}}$} associated with the estimates of $\mathbf{m_{jk}}$ for each observation considering the two models. We can observe that the overall variability of the DGLM-PPM (left panel) is smaller, specially when the time series goes through a change-point (observations indicated by a vertical line).

In short, our results show that the proposed model is more flexible than the traditional DGLM  while providing a richer inference without any loss of accuracy in the estimation of the dynamic parameters.
 In the next section, we display a real life example in which the DGLM-PPM, beyond providing the break-point analysis, also outperforms the DGLM both in out-of-sample prediction accuracy and in-sample measures of goodness-of-fit.

\section{Application}

\subsection{Model specification and comparison}

 In this section, we use the well-known Coal Mining data set to compare between the DGLM-PPM and the conventional DGLM. To make inferences on the partitions we used the MCMC scheme described previously. Since the application does not require the repeated use of the Gibbs Sampler algorithm as in the simulation experiment, we opted to generate a  longer single chain of size 45000. After a burn-in of 5000 and a lag of 10, 4000 posterior samples were obtained.
This final configuration was determined after some tests using different settings and with the help of the diagnostic tools for MCMC offered by the 'coda' package from R \cite{coda}. For this exercise, we considered Yao's block prior cohesion with a Beta prior distribution for the parameter $\pi$. The specification of the hyperparameters $r$ and $s$ can follow one of three approaches. First, we can assume no prior information about $\pi$. In this case, a non-informative  $\pi \sim \mathcal{B}(1,1)$ prior is a suitable choice and, according to Equation (\ref{priormeanB}), the prior expected number of change-points is equal to $\frac{n-1}{2}$. Another alternative is to attribute more weight to larger values of $\pi$ implying a high probability of occurrence for change-points. Such a prior would stimulate the formation of partitions with a larger number of blocks.   Conversely, a prior whose probability mass is concentrated in small values of $\pi$  implies that the modeler is expecting a small number of change-points in the process. In this example, we considered three prior specifications for $\pi$: $\mathcal{B}(1,1)$, $\mathcal{B}(1,10)$ and $\mathcal{B}(10,1)$. The results obtained with the DGLM-PPM under the three specifications were compared to those of the conventional DGLM both in-sample and out-of-sample.

In a Bayesian context, a popular in-sample method for model comparison is the Posterior Model Probability (PMP). Let $M_{i}$, $i=1,\cdots, N$ denote the $i$-th  model with N representing the total number of models in consideration. Then, the posterior probability of the model $M_{i}$ given the set $\bm{Y}$ can be defined as
\begin{equation}
    p(M_{i}\mid \bm{Y})=\frac{p(\bm{Y} \mid M_{i})p(M_{i})}{\sum_{i=1}^{N}p(\bm{Y} \mid M_{i})p(M_{i})},
    \label{PMP}
\end{equation}
where $p(M_{i})$ is the prior probability associated to $M_{i}$ and $p(\bm{Y} \mid M_{i})$ denotes the marginal distribution of $\bm{Y}$. In the context of the DGLM-PPM, given the partition $\rho$ and the discount factor $\delta$, $p(\bm{Y} \mid M_{i})$ can be obtained from Equation (\ref{preddglmppm}).

To access the predictive performance of each model we use the Mean Absolute Error (MAE) and the Mean Squared Error (MSE) defined, respectively, by
\begin{equation*}
    MAE=\frac{1}{n}\sum_{t=1}^{n}\abs{\hat{y_{t}}-y_{t}},
\end{equation*}
\begin{equation*}
   MSE=\frac{1}{n}\sum_{t=1}^{n}(\hat{y_{t}}-y_{t})^{2},
\end{equation*}
where $n$ denotes the size of the data set, $\hat{y_{t}}$ the one-step-ahead forecast for the instant $t$ and $y_{t}$ the actual realised value.

\subsection{Coal Mining data}

\begin{figure}[H]
    \centering
    \includegraphics[scale=0.5]{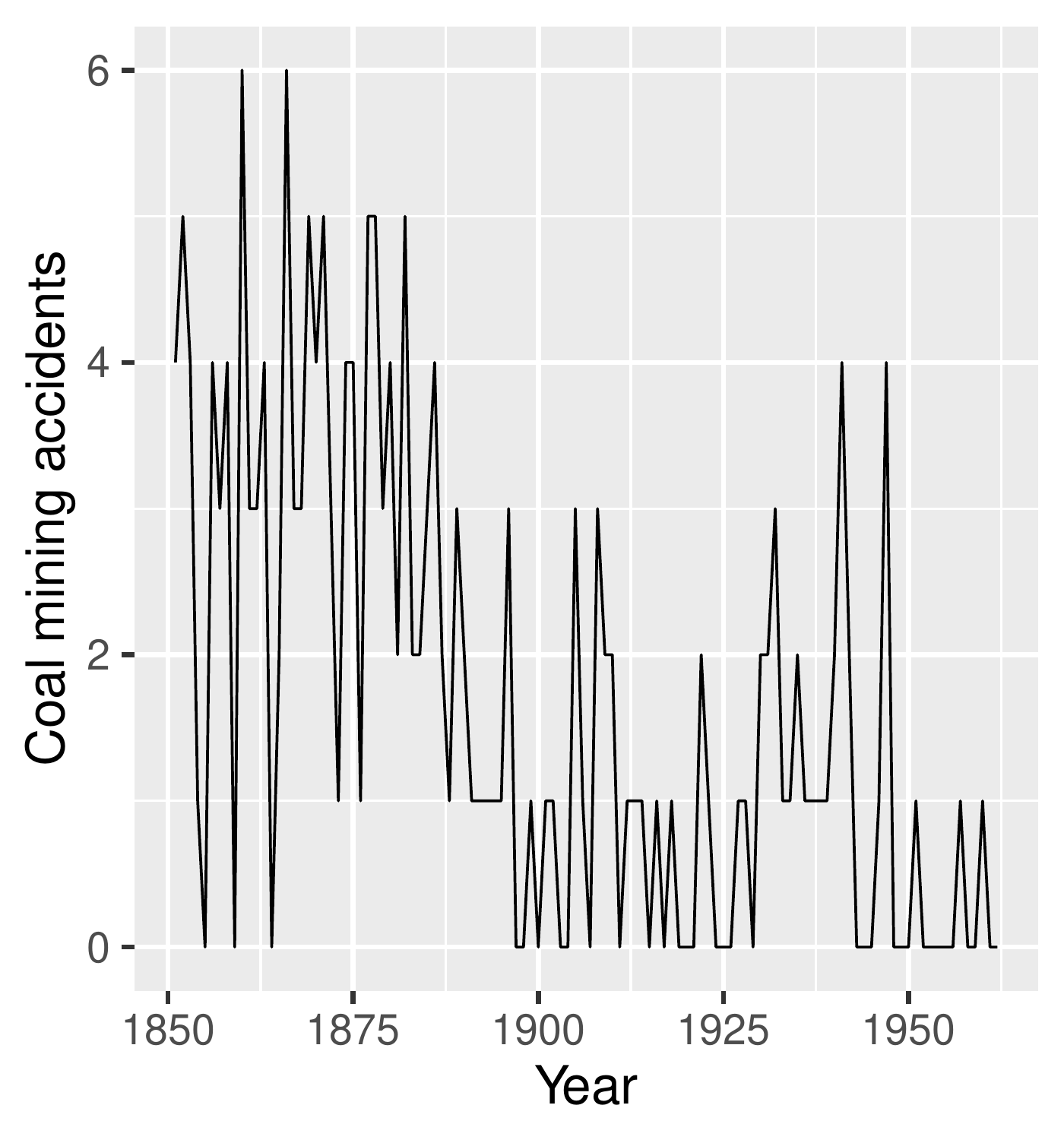}
    \caption{Annual number of coal mining disaster involving more than ten men  in the UK from 1851 to 1962.}
    \label{coal}
\end{figure}

 The coal mining data reports the annual number of coal mining disasters involving more than ten men in the UK from 1851 to 1962 totaling 112 observations. This time series has been extensively used in the literature of change-points in count data. Some examples include Raftry and Akman (1986) ~\cite{raftery86},  Carlin, Gelfan and Smith (1992) ~\cite{carlin92},  Santos, Franco and Gammerman (2010) ~\cite{santos2010}, Lai and Xing (2011) ~\cite{lai2011} and da Silva and da-Silva (2017) ~\cite{dasilva2017}. Figure (\ref{coal}) displays the coal mining disasters counts.

 Most works report two change-points for this time series -- the first one around the period that extends from 1885 to 1895 and the other around 1945. For this reason, we expect that the scenario in which $\pi \sim \mathcal{B}(10,1)$ should perform better.    A visual inspection of the time series plot does not indicate any signs of slope or seasonality. Thus, and since we are dealing with a time series of counts, a natural modeling choice is a Poisson Local Level Model as defined in Section 5. Also, no explanatory variables are available for this data set. The final model was specified with $G_{j}=F_{j}=1$ and initial state given by $(\bm{\theta}_{\rho}^{(j)} \mid D_{0})\sim(1,10)$.
 The specification of the variance of the evolution equation was made via a discount factor considering a $\mathcal{B}(1,1)$ prior distribution for $\delta$.

\subsection{Results}

Table (\ref{results}) summarizes the results obtained for the DGLM-PPM and the conventional DGLM. Along with the point estimates for the discount factors $\delta$ and the posterior number of blocks $B$, we also report their respective 95\% Credible Intervals (CI). The PMP was calculated according to equation (\ref{PMP}), assuming equal prior probability for each of the models considered.

\begin{table}[H]
\centering
\begin{tabular}{ccccc}
\hline
\multirow{2}{*}{\textbf{}}                                                                 & \multicolumn{3}{c}{\textbf{DGLM-PPM}}                                                                                                                                                       & \multirow{2}{*}{\textbf{\begin{tabular}[c]{@{}c@{}}Conventional \\ DGLM\end{tabular}}} \\ \cline{2-4}
                                                                                           & $\pi \sim \mathcal{B}(1,1)$                                  & $\pi \sim \mathcal{B}(1,10)$                                     & $\pi \sim \mathcal{B}(10,1)$                                        &                                                                                        \\ \hline \hline
\textbf{MAE}                                                                               & 1.017                                                        & 1.033                                                       & 1.0025                                                         & 1.045                                                                                  \\
\textbf{MSE}                                                                               & 1.74                                                         & 1.79                                                        & 1.70                                                           & 1.81                                                                                   \\
\textbf{PMP}                                                                               & 0.24                                                         & 0.087                                                       &0.59                                                          & 0.084                                                                                  \\
\textbf{\begin{tabular}[c]{@{}c@{}}Mean Discount Factor\\ (95\% CI)\end{tabular}}            & \begin{tabular}[c]{@{}c@{}}0.46\\ (0.077, 0.79)\end{tabular} & \begin{tabular}[c]{@{}c@{}}0.80\\ (0.70, 0.89)\end{tabular} & \begin{tabular}[c]{@{}c@{}}0.24\\ (0.0021,  0.51)\end{tabular} & \begin{tabular}[c]{@{}c@{}}0.82\\ (0.73, 0.90)\end{tabular}                            \\
\textbf{\begin{tabular}[c]{@{}c@{}}Mean Posterior Number \\ of blocks\\ (95\% CI)\end{tabular}} & \begin{tabular}[c]{@{}c@{}}29.90\\ (3, 68)\end{tabular}      & \begin{tabular}[c]{@{}c@{}}100\\ (83, 111)\end{tabular}     & \begin{tabular}[c]{@{}c@{}}11.50\\ (3, 23)\end{tabular}        & -                                                                                      \\ \hline
\end{tabular}
\caption{Summary of the results obtained with the DGLM-PPM and  DGLM}
\label{results}
\end{table}

We can observe that, in this exercise,  the DGLM-PPM out-performs the conventional DGLM both in forecast accuracy and goodness of fit regardless of the prior specification used for $\pi$. Also, as expected, among the specifications used for the  DGLM-PPM, there is a clear advantage for the scenario in which  $\pi \sim \mathcal{B}(10,1)$. This specification presented a PMP of 59\%, making it the most suitable choice for the Coal Mining Disasters counts and reflecting the fact that this data set has few change-points. It is noteworthy that the DGLM-PPM is sensitive to the prior specification of $\pi$. Thus, the modeler must be careful to choose an informative prior that mirrors the visual information provided by the time series plot.

\begin{figure}[H]
\centering
\caption{Coal mining application: (a) Discount Factors; (b) Posterior number of blocks in the partition.}
\begin{subfigure}{.5\textwidth}
  \centering
  \includegraphics[scale=0.5]{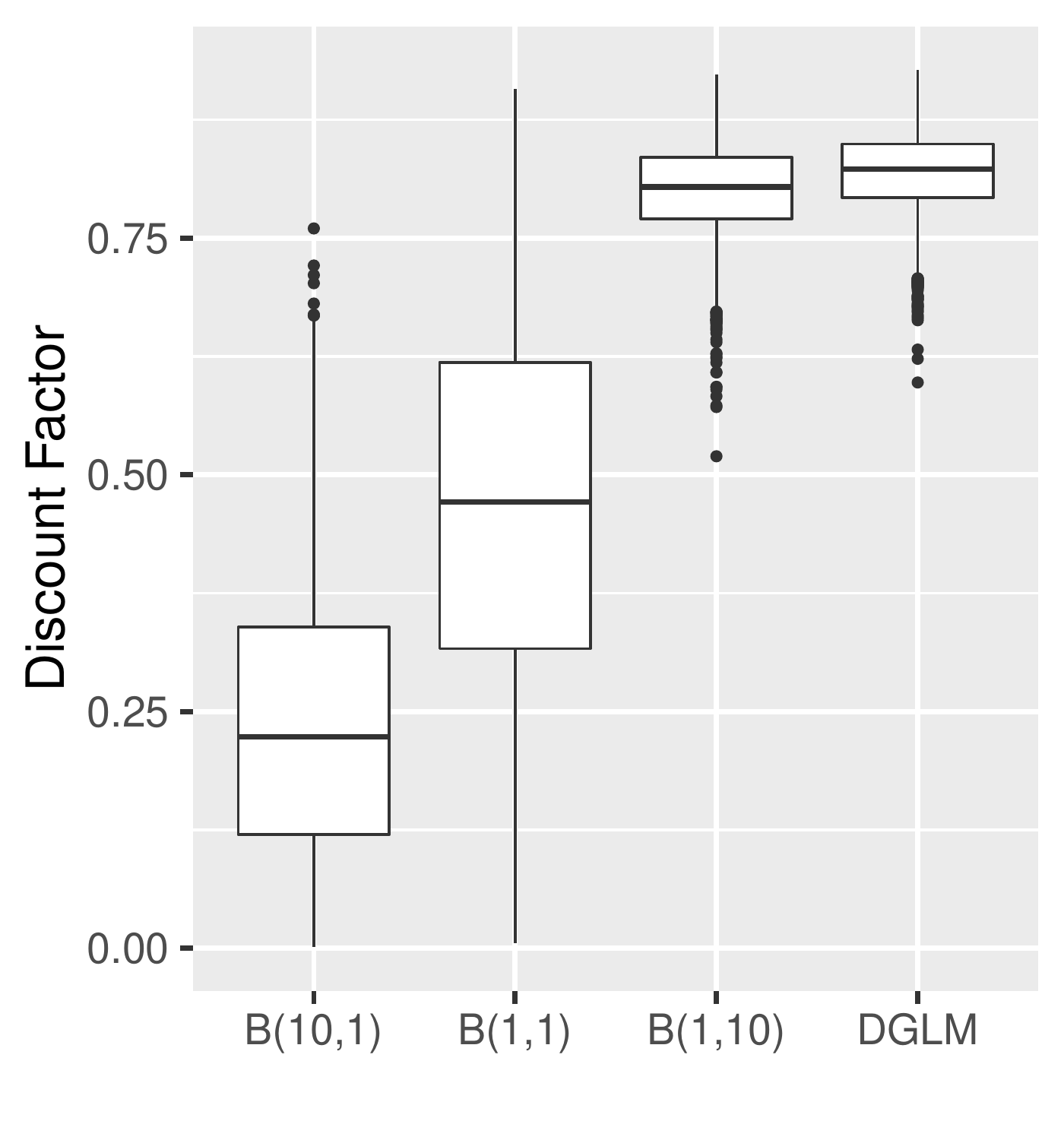}
 \caption{}
  \label{discf}
\end{subfigure}%
\begin{subfigure}{.5\textwidth}
  \centering
  \includegraphics[scale=0.5]{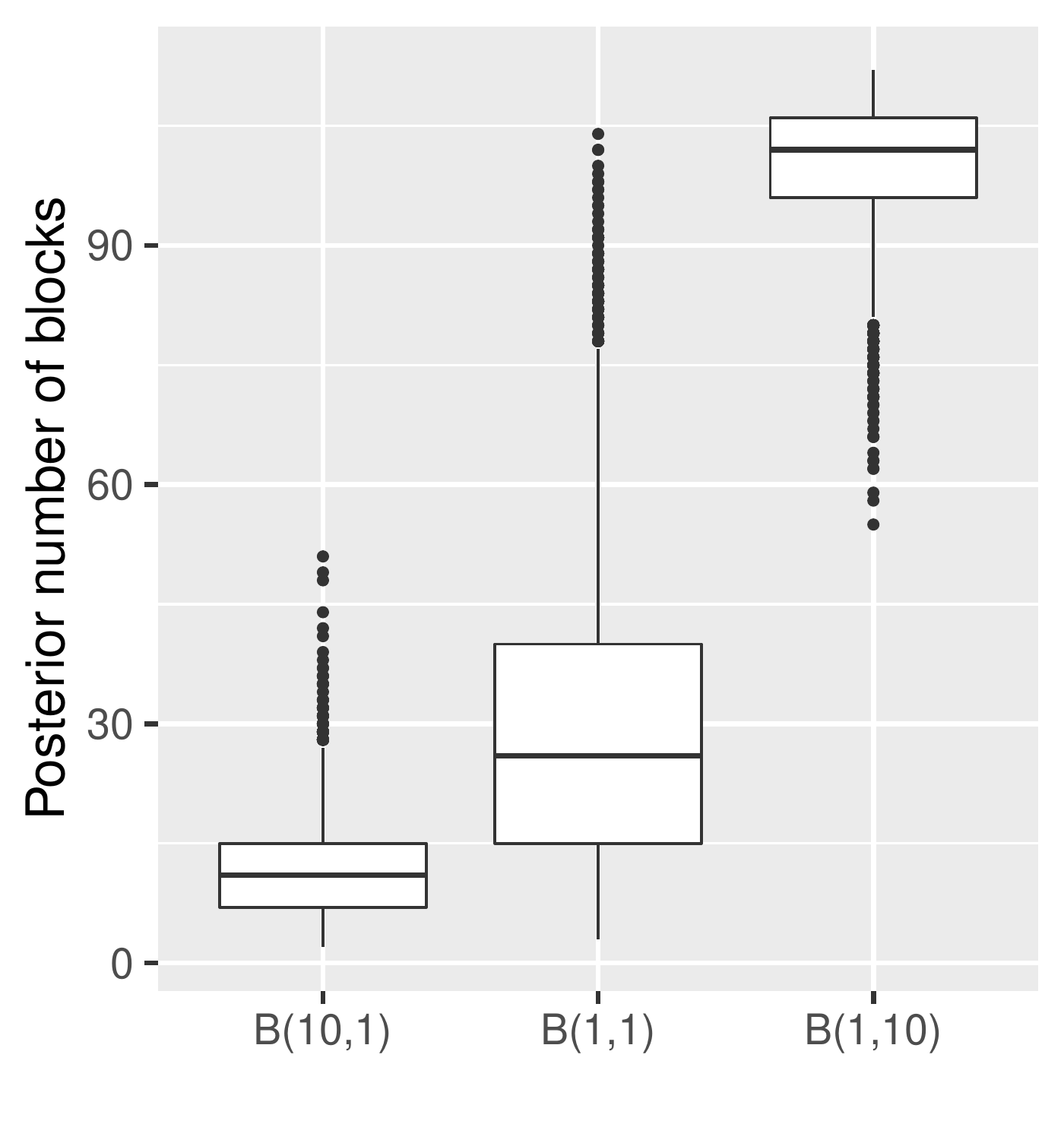}
 \caption{}
  \label{nbl}
\end{subfigure}
\end{figure}

The results concerning the discount factors and posterior number of blocks are also in accordance with  the simulation study and the discussions carried out in the last sections. The discount factors associated with the DGLM-PPM are smaller than the one estimated for the conventional DGLM in, at least, two of the three scenarios tested. We also observe that larger values of $\delta$ are related to a larger number of change-points. The scenario considering  $\pi \sim \mathcal{B}(1,10)$, for instance, generates partitions of mean size $\hat{B}=100$ which is consistent with a situation of a highly unstable process that seems  not be the case for the Coal Mining accidents data. On the other hand, the scenario described by the prior $\pi \sim \mathcal{B}(10,1)$ produces much smaller partitions and looks more likely. The boxplots in figures (\ref{discf}) and (\ref{nbl}) provide a visualization of the empirical distributions of the posterior samples obtained for $\delta$ and $B$, respectively.

\begin{figure}[H]
\centering
\includegraphics[scale=0.5]{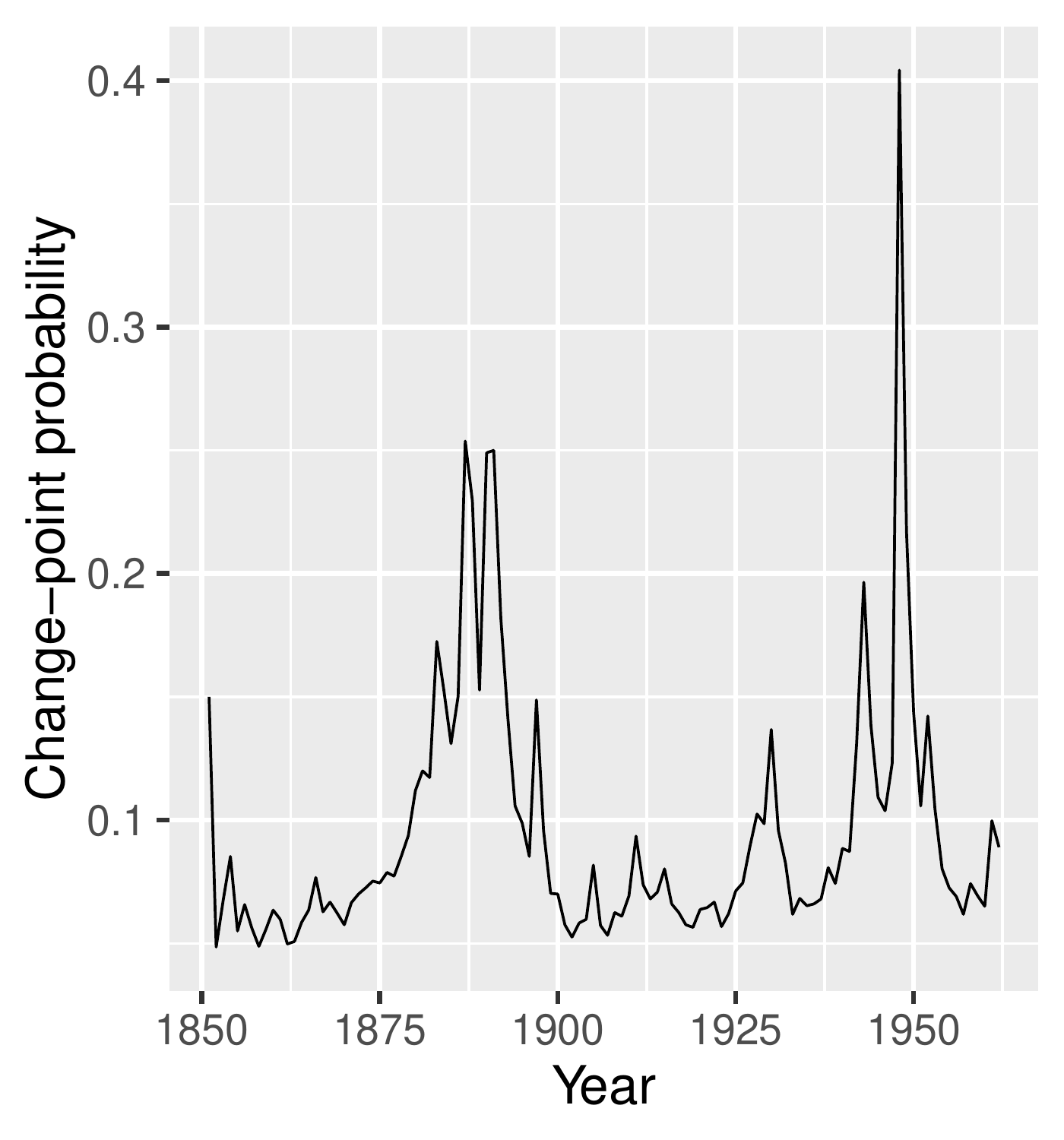}
\caption{Change-point probabilities for the coal mining accidentes time series using the DGLM-PPM with $\pi \sim \mathcal{B}(10,1)$.}
\label{probchange}
\end{figure}

Since the scenario considering $\pi \sim \mathcal{B}(10,1)$ provided the best fit according to all the criteria evaluated, the following analysis will consider only this model specification. One of the main advantages of the DGLM via PPM is that it allows online inference as in the DGLM class along with estimates of the change-point probabilities for every observation in the time series under study. Figure (\ref{probchange}) displays the probability of change estimated for each observation of the coal mining accident data set using the DGLM-PPM. Two significant peaks can be observed: the first, smaller, around 1890, and the second round 1947. Our results are consistent with the findings of Lai \& Xing (2011). They report change-point probabilities in the order of 40\% for the points around 1950 and of $~20\%$ for the observations near 1890. Similar results were obtained by da-Silva and da Silva (2017) using a Chopin Filter based model.

\begin{figure}[H]
\centering
\includegraphics[scale=0.5]{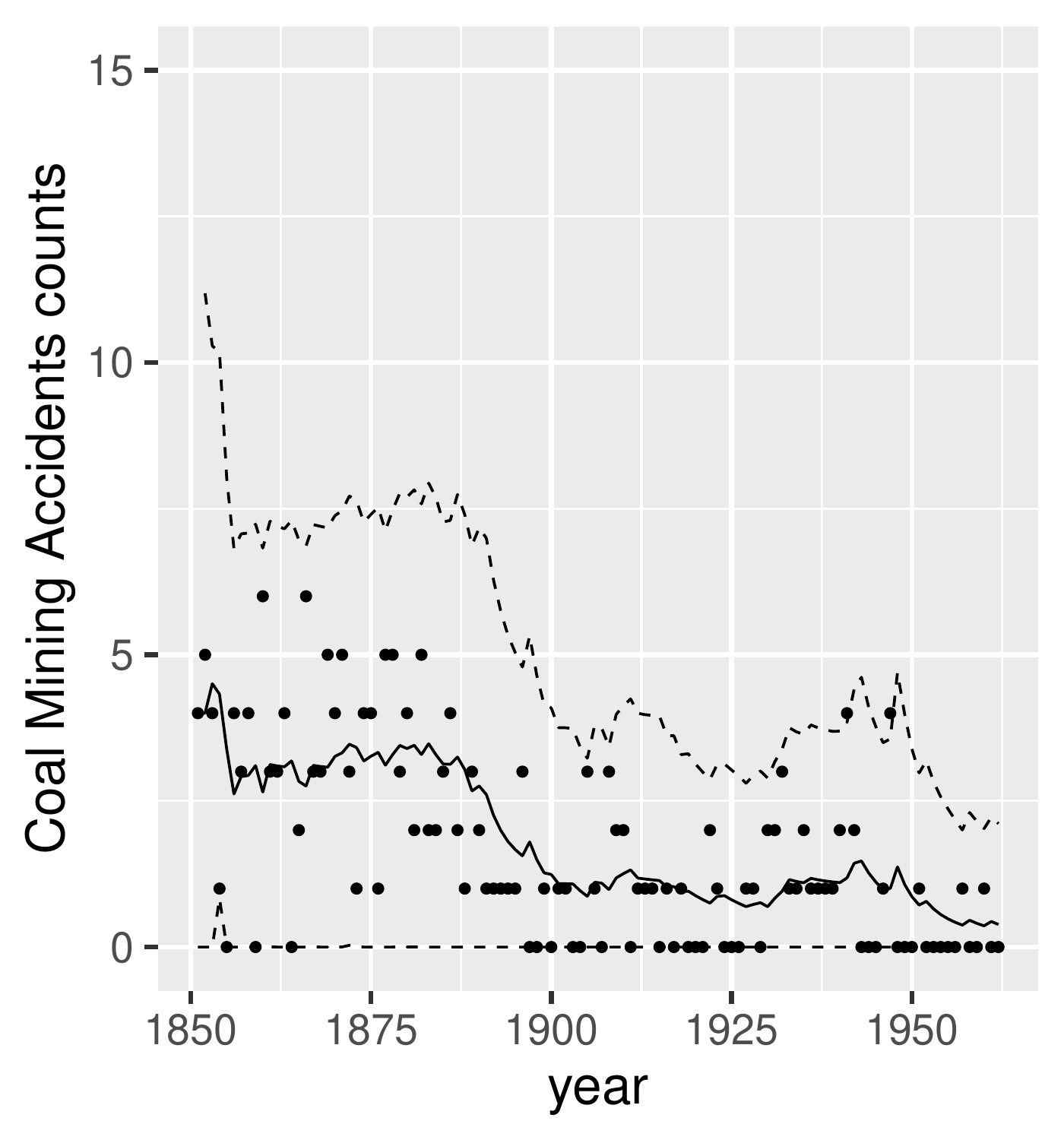}
\caption{One-step-ahead forecasts for the coal mining accidents time series using the DGLM-PPM with $\pi \sim \mathcal{B}(10,1)$. The solid black line represents the predictions; the dashed red lines indicate the 95\% CI}
\label{pred}
\end{figure}

The DGLM-PPM also allows the computation of predictions. Figure (\ref{pred}) displays the one-step-ahead forecasts along with a 95\% CI. The point and interval estimates were calculated as the expected values and quantiles of the predictive distribution (\ref{preddglmppm}). Observe that the predictions line (in black) follows the data points smoothly. Also, all the data points are within the boundaries set by the credible intervals (dashed red lines), indicating a good fit.

\section{Conclusions}

Intervention analysis is a significant concern in time series analysis. In this paper, we presented the DGLM-PPM -- a new class of models that incorporates the Product Partitions Models of Barry \& Hartigan (1990) ~\cite{hartigan90} into the theory of Dynamic Generalized Linear Models. We believe our model presents several improvements over the conventional DGLM. First, it allows the identification of structural breaks by assigning to each observation in the time series a posterior probability that it is a change-point. Unlike many models used to detect change-points, the DGLM-PPM does not require any prior knowledge regarding the observation wherein the regime switch happens, which can be an essential advantage in many situations. Also, just like any other model in the DGLM class, the DGLM-PPM permits the estimation of filtered and smoothed states, allowing for k-step-ahead forecasts and retrospective analysis.

Another key feature of the DGLM-PPM class is the possibility of making inferences on the optimal block structure for a determined data set. Samples of the partition can be obtained in an efficient manner using a Gibbs Sampler scheme proposed by Barry \& Hartigan (1993) ~\cite{barry93}. Within this Gibbs scheme, we included an ARMS ~\cite{gilks95} step in order to sample from the posterior distribution of the discount factor.

To evaluate the properties of the DGLM-PPM, we conducted a simulation experiment. From this study, the main conclusions were: i) as expected, the model assigns change-point probabilities very close to one to those observations where artificial jumps were created; ii) the discount factor associated to the DGLM-PPM are smaller than those estimated for the conventional DGLM. This outcome is a result of the block structure induced by the PPM; iii) the median relative biases of the posterior mean of the state vector is very small, showing good adjustment of the DGLM-PPM. We also observe that, immediately after a change-point, there is a learning process in which the large estimation biases caused by the structural break get progressively smaller as the observations of the time series are processed. This effect, however, is much faster in the DGLM-PPM in comparison to the traditional DGLM; iv) The estimated posterior number of blocks in the partition is higher than we would have expected. This is probably due to the prior cohesion chosen and the variability in the generation of the artificial data.


The DGLM-PPM was applied to the well-known UK coal mining time series. Yao's prior cohesion was assumed with three different prior for the parameter $\pi$. We compared the DGLM-PPM under the specifications considered to the conventional DGLM using measures of in-sample and out-of-sample performance. The results showed that the DGLM-PPM outperformed the conventional DGLM regardless the criteria used to compare the models. We also verified that the DGLM-PPM detected possible change-points that are in line with the findings of other works in the literature.

Our results indicate that the methodology developed here can be a useful tool to detect change-points in time series while also improving the online and retrospective inferences provided by the conventional DGLM. Future works include a thoroughly conducted study of the effect of different prior cohesion in the results obtained with the DGLM-PPM and the incorporation of a stochastic evolution within blocks. The authors also envision the utilization of the DGLM-PPM as a support tool for piecewise regression.

\bibliographystyle{unsrt}


\end{document}